\documentclass[nofootinbib, aps,11pt,preprintnumbers]{revtex4-1}

\usepackage{graphicx}
\usepackage{cancel}
\usepackage{amssymb}
\usepackage{textcomp}
\usepackage{amsmath}
\usepackage{bm}
\usepackage{times}
\usepackage{epsfig}
\usepackage{color}
\def\beq{\begin{equation}}
\def\eeq{\end{equation}}
\def\bea{\begin{eqnarray}}
\def\eea{\end{eqnarray}}

\makeatletter
\def\blfootnote{\xdef\@thefnmark{}\@footnotetext}
\makeatother

\begin{document}
\title{\Large  {{\bf{LSP Squark Decays at the LHC \\
and the Neutrino Mass Hierarchy}}}}
\author{{Zachary Marshall${}^{1}$, Burt A.~Ovrut${}^{2}$, Austin Purves${}^{2}$ and Sogee Spinner${}^{2}$} \\[5mm]
    {\it  ${}^{1}$ Physics Division, Lawrence Berkeley National Laboratory} \\
   {\it Berkeley, CA 94704}\\[4mm]
    {\it  ${}^{2}$ Department of Physics, University of Pennsylvania} \\
   {\it Philadelphia, PA 19104--6396}\\[4mm]
}
\date{\today}

\blfootnote{zlmarshall@lbl.gov, ~ovrut@elcapitan.hep.upenn.edu,  ~apurves@sas.upenn.edu, ~sogee@sas.upenn.edu}
\begin{abstract}

The existence of $R$-parity in supersymmetric models can be naturally explained as being a discrete subgroup of gauged baryon minus lepton number ($B-L$). The most minimal supersymmetric $B-L$ model triggers spontaneous $R$-parity violation, while remaining consistent with proton stability. This model is well-motivated by string theory and makes several interesting, testable predictions. Furthermore, $R$-parity violation contributes to neutrino masses, thereby connecting the neutrino sector to the decay of the lightest supersymmetric particle (LSP). This paper analyzes the decays of third generation squark LSPs into a quark and a lepton. In certain cases, the branching ratios into charged leptons reveal information about the neutrino mass hierarchy, a current goal of experimental neutrino physics, as well as the $\theta_{23}$ neutrino mixing angle. Furthermore, optimization of leptoquark searches for this scenario is discussed. Using currently available data, the lower bounds on the third generation squarks are computed.
\end{abstract}

\maketitle
%

\section{Introduction}

The upgrade to the Large Hadron Collider (LHC) will soon be completed, providing us with an exciting opportunity to probe the next energy frontier. Among the many candidates for new physics in that frontier, supersymmetry (SUSY) stands out as a rich and compelling framework. SUSY not only addresses the gauge hierarchy problem, a puzzle that has driven many model building efforts over several decades, but can also speak to other outstanding issues in the standard model (SM). This includes dark matter and a mechanism for radiative electroweak symmetry breaking. As we wait for the next LHC run to begin, the interim is a good period to reconsider the phenomenology of low energy supersymmetric models. Among other things, it is of interest to investigate if they can yield any signals that have not yet been seriously considered, especially in well-motivated alternatives to the $R$-parity conserving minimal supersymmetric standard model (MSSM).

Despite their theoretically pleasing aspects, generic SUSY particle physics models potentially have  a serious problem regarding proton decay. This follows from the fact that the most general MSSM superpotential allows for baryon and lepton number violating terms at tree level and, therefore, rapid proton decay. The typical, yet \textit{ad hoc}, solution is to impose $R$-parity, $R_P = (-1)^{3 (B-L) +2s}$ where $s$ is the spin of the particle. This discrete symmetry forbids violation of baryon number ($B$) minus lepton number ($L$) by one unit. Accepting $R$-parity conservation, however, severely narrows one's view of the SUSY phenomenological landscape. This is because the lightest supersymmetric particle (LSP) in $R$-parity conserving theories is stable and, therefore, must be neutral due to cosmological considerations.

Perhaps the most appealing candidates for a deeper origin for $R$-parity, models with gauged $U(1)_{B-L}$, are based on the observation that $R$-parity is a discrete subgroup of $U(1)_{B-L}$. In such models, $R$-parity is a good symmetry as long as $U(1)_{B-L}$ is. However, once $U(1)_{B-L}$ is broken, the $B-L$ number of the field that breaks $U(1)_{B-L}$ determines the fate of $R$-parity: an even $B-L$ field leads to automatic $R$-parity conservation (RPC)~\cite{Mohapatra:1986su, Krauss:1988zc, Font:1989ai, Martin:1992mq} (for more recent studies see ~\cite{Aulakh:1999cd, Aulakh:2000sn, Babu:2008ep, Feldman:2011ms, FileviezPerez:2011dg}) , while an odd $B-L$ field triggers spontaneous $R$-parity violation (RPV)~\cite{Aulakh:1982yn, Hayashi:1984rd, Mohapatra:1986aw, Masiero:1990uj}\footnote
{
	See also recent studies of explicit $R$-parity violation assuming minimal flavor violation~\cite{oai:arXiv.org:0710.3129, oai:arXiv.org:1111.1239}.
}.
Typically, spontaneous $R$-parity violation is safe in the sense that only lepton number violation is generated at tree level, leaving the proton as stable as it would be with RPC.

As one might expect, the approach in these early $B-L$ studies was to introduce a new ``Higgs'' sector (that is, superfields with a $B-L$ charge) with which to spontaneously break the $B-L$ symmetry. However, the $B-L$ anomaly cancellation conditions provide a subtle, and more minimal, alternative to this approach. Note that the three generations of right-handed neutrino superfields required to cancel these anomalies contain right-handed sneutrinos. Remarkably, the right-handed sneutrinos have the correct quantum numbers to spontaneously break $B-L$ in a phenomenologically acceptable way. Specifically, they are neutral under the SM, carry no baryon number and, of course, have a $B-L$ charge of one. Therefore, anomaly cancellation defines the most minimal $B-L$ extension of the MSSM. This model has exactly the MSSM particle content plus three generations of right-handed neutrino supermultiplets, and it does {\it not} require a new Higgs sector. This minimal $B-L$ theory was proposed in~\cite{FileviezPerez:2008sx, Barger:2008wn, FP:2009gr, Everett:2009vy}, arguing for it's appeal from a ``bottom up'' point of view.\footnote
{
Such a minimal model was outlined as a possible low energy manifestation of $E_6$ grand unified theory (GUT) models in~\cite{Mohapatra:1986aw}.
}
The same theory was found from a ``top down'' approach within the context of a class of vacua of heterotic $M$-theory \cite{Evans:1986ada, Lukas:1998yy, Braun:2005ux, Braun:2005nv, Braun:2006ae, Ambroso:2009jd}. Due to the odd $B-L$ charge of the sneutrino, the minimal $B-L$ model must always spontaneously break $R$-parity. However, because the right-handed sneutrino has no baryon number, it's vacuum expectation value (VEV) does not introduce proton decay at tree level. 
In addition, this model has several potentially testable and interesting predictions:
\begin{itemize}

\item $R$-parity violation is manifest though lepton number violating operators, which could lead to lepton number violating signatures at the LHC, \textit{e.g.}~\cite{FileviezPerez:2012mj, Perez:2013kla}.

\item The existence of two neutral light fermions (sterile neutrinos), in addition to the usual three neutrinos~\cite{Mohapatra:1986aw, Ghosh:2010hy, Barger:2010iv}. These may play a role in cosmology~\cite{Ghosh:2010hy, Ade:2013zuv, Perez:2013kla}.

\item A $B-L$ neutral gauge boson, $Z'$, whose mass is proportional to the soft mass of the right-handed sneutrino. This gauge boson must be at the TeV scale and, therefore, detectable at the LHC.

\item The right-handed sneutrino VEV directly links the neutrino sector to lepton number violation by one unit. This generates tree-level Majorana contributions to the neutrino masses.
\end{itemize} 

This last statement is significant, since it specifies the size of the RPV. It follows from the upper bound placed on this contribution by the neutrino masses that the RPV is only relevant for the decay of the LSP, which would otherwise be stable under RPC. All other SUSY processes will effectively be $R$-parity conserving. The last bullet point is also crucial because it relates neutrino masses to collider physics through $R$-parity violation, an exciting synergy. It suggests that one may be able to infer information about the neutrino sector from LSP decays. Finally, it is worthwhile to note that despite RPV, a gravitino LSP, while unstable, may live long enough to be the dark matter of the universe~~\cite{Borgani:1996ag, Takayama:2000uz, Buchmuller:2007ui}.

This model of spontaneous RPV is, therefore, a well-motivated alternative to RPC. As with all SUSY models, its phenomenology will be highly depended on the choice of the LSP\footnote
{
While the complete model would include a gravitino LSP as the dark matter of the universe, throughout this paper we shall use LSP to refer to the lightest supersymmetric particle {\it relevant for collider physics}.
}. $R$-parity violation plays an important role from this perspective because it allows the LSP to decay. This liberates the LSP to be any superpartner, including those that have color and charge. One example, of this type, is a charged slepton LSP. However, this will decay like a charged Higgs, an element that already exists in the MSSM. Squark LSPs, on the other hand, offer an opportunity for a whole new set of signals since they act as leptoquarks; that is, scalar particles that are pair produced and decay into a quark and a lepton. Among the squarks, the third generation is perhaps the most interesting LSP candidate since these are generally expected to have the lowest masses due to renormalization group effects, \textit{e.g.}~\cite{Martin:1997ns}. Furthermore, since the lower generations must be fairly degenerate due to the SUSY flavor problem, they would be produced more readily and, therefore, have stronger bounds. Finally, stops are the most engaging of all the squarks because of their substantial radiative contribution to the Higgs mass and the role they play as a measure of fine-tuning in SUSY; that is, the little hierarchy problem.

Motivated by this discussion, this paper extends the study of our earlier paper~\cite{Marshall:2014kea}, by analyzing the prompt decays of third generation squark LSPs within the context of a minimal $B-L$ extension of the MSSM. One of the aims of this paper is to highlight the relationship between stop and sbottom LSP decays and the neutrino sector. Especially striking is the fact that one may infer information about the neutrino mass hierarchy from the $R$-parity violating LSP decays. Just as important are the leptoquark signals, which are typically not associated with SUSY. Experimentally, they have not yet been analyzed with data from the latest LHC run. As we will show in this paper, the leptoquark searches that have previously been conducted allow stop LSP masses as low as 420 GeV and sbottom LSP masses as low as 500 GeV.

The rest of this paper is organized as follows. Section~\ref{the.model} introduces the details of the model as well as specifying the $R$-parity violating sector. The consequences in terms of $R$-parity violation are discussed in Section~\ref{RPV} and their influence on neutrino masses are illustrated in Section~\ref{sec.nu.mass}. Section~\ref{stop} contains the results for both the stops and sbottoms, including lower bounds and the connection between squark decays and the neutrino sector. This connection is explored through a numerical scan, but the results can be understood analytically, an is done in Section~\ref{disc}. Section~\ref{disc} also attempts to frame the results in terms of a bigger picture, investigating how this scenario can be distinguished from scenarios with similar signatures. Finally, Section~\ref{out} summarizes our results. Throughout this work, many references will be made to technical calculations discussed in Appendix~\ref{nu}, making this a potentially important section for the reader. The remaining three Appendices, B, C and D, briefly discuss the chargino sector, the third generation squark sector and the Feynman rules used in the calculations of the squark decays.

\section{The Minimal SUSY B-L and Spontaneous R-parity Violation}
\label{the.model}
There are several possible minimal $B-L$ extensions of the MSSM of the form $SU(3)_C \otimes SU(2)_L \otimes U(1) \otimes U(1)'$, characterized by different choices of the two $U(1)$ factors. If these are remnants of a GUT theory, such as SO(10), then these possibilities are all physically equivalent, but will be characterized by different kinetic mixing between the two $U(1)$ factors. Among these possibilities, as shown in \cite{Ovrut:2012wg}, there is a unique choice that will have vanishing kinetic mixing--not only at the GUT scale, but at any lower scale. This choice of $U(1)$ factors is $U(1)_{3R} \times U(1)_{B-L}$, where $U(1)_{3R}$ is the third component of right-handed isospin.   The fact that this basis has no kinetic mixing greatly simplifies the present analysis. Therefore, in this paper, we proceed using the specific minimal extension gauge group
\begin{equation}
	SU(3)_C \otimes SU(2)_L \otimes U(1)_{3R} \otimes U(1)_{B-L} \ .
\end{equation} 
We will comment later in the paper on how our results apply to the other similar extensions. The gauge structure in this case is such that the hypercharge, $Y$, is related to the $B-L$ and third component of right-handed isospin charges by
\begin{equation}
	Y = I_{3R} + \frac{B-L}{2},
\end{equation}
analogous to the relationship between the electric charge, hypercharge and third component of left-handed isospin in the SM. 

The matter content and its $SU(3)_C \otimes SU(2)_L \otimes U(1)_{3R} \otimes U(1)_{B-L}$ charges is given by three copies of
\begin{equation}
	Q \sim (\textbf{3}, \textbf{2},0,1/3), \  \,  u^c \sim (\bar{\textbf{3}},\textbf{1},-1/2,-1/3), \  \,  {d}^c \sim (\bar{\textbf{3}},\textbf{1},1/2,-1/3),
\end{equation}
\begin{equation}
	{L} \sim (\textbf{1},\textbf{2},0,-1), \  \,  {e}^c \sim (\textbf{1},\textbf{1},1/2,1), \  \,   {\nu}^c \sim (\textbf{1},\textbf{1},-1/2,1), \\
\end{equation}
while the MSSM Higgs sector is
\begin{displaymath}
	{H}_u \sim (\textbf{1},\textbf{2},1/2,0) , \ \  {H}_d \sim (\textbf{1},\textbf{2},-1/2,0).
\end{displaymath}
The superpotential is similar to that of the MSSM but contains an additional Yukawa coupling to the right-handed neutrino superfield
\begin{equation}
\label{W}
	W =
	Y_u {Q} {H}_u {u}^c - Y_d {Q} {H}_d {d}^c
	- Y_e {L} {H}_d {e}^c + Y_\nu {L} {H}_u {\nu}^c
	+ \mu  H_u  H_d,
\end{equation}
where the Yukawa couplings are three-by-three matrices in family space and are in general complex. The soft SUSY breaking Lagrangian is
\begin{align}
	\begin{split}
	-\mathcal{L} _{\rm soft}= &
		\, m_{\tilde \nu^c}^2 |\tilde \nu^c|^2 + m_{\tilde L}^2 |\tilde L|^2 + m_{H_u}^2 |H_u|^2  + m_{H_d}^2 |H_d|^2
\\
		& +
		\left(
			M_R \tilde W_R^2 + M_2 \tilde W^2 + M_{BL} \tilde B'^2 + M_3 \tilde g^2 + a_\nu \tilde L H_u \tilde \nu^c
			+ b H_u H_d+ \text{h.c.}
		\right)+ \cdots,
	\end{split}
\end{align}
where the ellipses refer to terms which also exist in the MSSM and are not crucial here. The fields $\tilde W_R, \tilde W, \tilde B'$ and $\tilde g$ are the fermion superpartners of the third component of right-handed isospin, left-handed isospin, $B-L$ and color gauge bosons respectively. The $a_\nu$ is the soft trilinear analogue of $Y_\nu$ and is, therefore, also a three-by-three matrix in family space. The superpotential and Lagrangian are valid in the energy regime between the GUT scale and the TeV scale. Here we continue by analyzing physics at the TeV scale.
 
The notation for the VEVs of the fields phenomenologically allowed to acquire sizable VEVs is
\begin{equation}
\label{vevs}
	\left< \tilde \nu^c_3 \right> \equiv \frac{1}{\sqrt 2} v_R, \ \ \left<\tilde \nu_i\right> \equiv \frac{1}{\sqrt 2} {v_L}_i, \ \
	\left< H_u^0\right> \equiv \frac{1}{\sqrt 2}v_u, \ \ \left< H_d^0\right> \equiv \frac{1}{\sqrt 2}v_d,
\end{equation}
where $i=1,2,3$ is the generational index and $\tan \beta \equiv v_u/v_d$. The generation of a right-handed sneutrino superfield is not identifiable through its interactions, unlike a left-handed electron neutrino which couples to the electron through the $SU(2)_L$ gauge interactions. As a result, there is freedom to rotate the right-handed neutrino fields into any basis and specifically to a basis in which only one generation of right-handed sneutrino acquires a VEV. Here this will be chosen, without loss of generality, to be the third generation. Electroweak symmetry breaking will induce VEVs in the remaining two right-handed sneutrino generations. However, these will be on the order of the neutrino masses and, therefore, are neglibible. Note that ${v_L}_i$ is in general complex.

Substituting the VEVs from Eq.~(\ref{vevs}) into the $F$-term, $D$-term and soft potentials yields
\begin{align}
\label{F}
\begin{split}
\langle V_F\rangle &=
	\frac{1}{2}|\mu|^2 v^2+\frac{1}{4}|{Y_\nu}_{i3} {v_L}_i|^2v_R^2 + \frac{1}{4}|{Y_\nu}_{i3}|^2v_u^2v_R^2 + \frac{1}{4} |{Y_\nu}_{ij} {v_L}_i|^2v_u^2
\\
	& \quad  -\frac{1}{2\sqrt 2}(\mu {Y_\nu}_{i3} v_d {v_L}_i v_R+\text{h.c.}),
\end{split}
\\
\langle V_D\rangle&=\frac{g_2^2}{32}(v_u^2-v_d^2- |{v_L}_i|^2)^2 + \frac{g_{BL}^2}{32}(v_R^2- |{v_L}_i|^2)^2 + \frac{g_R^2}{32}(v_u^2-v_d^2-v_R^2)^2,
\\
\label{soft}
\langle V_{\mbox{soft}}\rangle&=\frac{1}{2}m^2_{\tilde L_{i}} \left| {v_L}_i \right|^2 + \frac{1}{2}m^2_{\tilde \nu_3^c}v_R^2 + \frac{1}{2}m_{H_u}^2v_u^2 + \frac{1}{2}m_{H_d}^2v_d^2 + b \, v_d v_u + \frac{1}{2\sqrt2}(a_{\nu_{i3}}v_u {v_L}_i v_R+\mbox{h.c.}),
\end{align}
where repeated generational indices are summed and $g_R, g_2$ and $g_{BL}$ are the third component of right-handed isospin, left-handed isospin and $B-L$ gauge couplings respectively.

Equations~(\ref{F})-(\ref{soft}) can be simplified by considering some general phenomenological features of this model. For example, neutrino masses are roughly proportional to the ${Y_\nu}_{ij}$ and ${v_L}_i$ parameters and, hence, ${Y_\nu}_{ij} \ll 1$ and ${v_L}_i\ll v_{u,d}, v_R$. With this in mind, the complete potential energy has the following minimization conditions:
\begin{align}
v_R^2=&\frac{-8m^2_{\tilde \nu_{3}^c}  + g_R^2\left(v_u^2 - v_d^2 \right)}{g_R^2+g_{BL}^2}\\
\frac{1}{8}(g_2^2+g_R^2)v^2=&-|\mu|^2+\frac{M_{H_u}^2\tan^2\beta-M_{H_d}^2}{1-\tan^2\beta}\\
\frac{2b}{\sin2\beta}=&2|\mu|^2+m_{H_u}^2+m_{H_d}^2
\end{align}
\begin{align}
{v_L}_i=&\frac{\frac{v_R}{\sqrt 2}(Y_{\nu_{i3}}^* \mu v_d-a_{\nu_{i3}}^* v_u)}{m_{\tilde L_{i}}^2-\frac{g_2^2}{8}(v_u^2-v_d^2)-\frac{g_{BL}^2}{8}v_R^2}
\end{align}
where $v^2 = v_d^2 + v_u^2$ and 
\begin{eqnarray}
	M_{H_u}^2&\equiv&m_{H_u}^2-\frac{1}{8}g_R^2v_R^2\\
	M_{H_d}^2&\equiv&m_{H_d}^2+\frac{1}{8}g_R^2v_R^2 \ .
\end{eqnarray}
These conditions necessarily mean that the soft mass of the sneutrino that acquires a VEV, the third generation here, must have a tachyonic soft mass. Radiative mechanisms for achieving such a mass have been discussed in references~\cite{Ambroso:2009jd, Ambroso:2009sc, Ambroso:2010pe}.

Prior to electroweak symmetry breaking, $B-L$ breaking leaves one linear combination of the third component of right-handed isospin and $B-L$ gauge bosons massless-- the hypercharge gauge boson. The other linear combination, $Z_R$, becomes massive. Including electroweak symmetry breaking effects, the mass of $Z_R$ is
\begin{equation}
	M_{Z_R}^2 \simeq \frac{1}{4} \left(g_R^2 + g_{BL}^2\right) v_R^2
	\left(
		1 + \frac{g_R^4}{\left(g_R^2+g_{BL}^2\right)^2} \frac{v^2}{v_R^2}
	\right).
\end{equation}
See reference~\cite{Everett:2009vy} for more details. Current bounds on $M_{Z_R}$ are at around 2.5 TeV~\cite{ATLAS:2013jma,CMS:2013qca}.

\section{R-parity Violation}
\label{RPV}
$R$-parity violation in this model is best parameterized by the two flavorful parameters-- ${v_L}_i$ and
\begin{equation}
	\epsilon_i  \equiv \frac{1}{\sqrt 2} {Y_\nu}_{i3} v_R \ .
\end{equation}
The superpotential expanded around the vacuum now contains the $R$-parity violating terms
\begin{equation}
	\label{W.brpv}
	W \supset \epsilon_i \,  L_i \,  H_u - \frac{1}{\sqrt 2 }{Y_e}_i \, {v_L}_i \,  H_d^- \,   e^c_i \ ,
\end{equation}
which is similar to the so-called bilinear RPV scenario~\cite{Hirsch:2000ef}. In addition, the Lagrangian contains various other bilinear terms, generated by ${v_L}_i$ and $v_R$, from the super-covariant derivative:
\begin{align}
	\label{L.n}
	\mathcal{L} \supset
	- \frac{1}{2}{v_L}_i^* \left[ g_2 \left(\sqrt 2 \, e_i \tilde W^+ 
	+  \nu_i \tilde W^0\right) - g_{BL} \nu_i \tilde B' \right]
	-\frac{1}{2} v_R \left[-g_R \nu_3^c \tilde W_R + g_{BL} \nu_3^c \tilde B' \right]+ \text{h.c.}
\end{align}

The results and analysis in the paper will be carried out using the Lagrangian based on Eqs.~(\ref{W.brpv}) and (\ref{L.n}). However, it is worthwhile to note that it is sometimes useful to rotate away the $\epsilon_{i}$ term in favor of the so-called trilinear $R$-parity violating terms. This is true when comparing to given bounds on various low-energy constraints on RPV, such as lepton number violating processes, and it makes approximating decays widths more straightforward. An example of each of these will be given in this section. Rotating $\epsilon_i$ away generates the following terms in the superpotential:
\begin{equation}
\label{eq:TRPV}
	W_{TRPV} = \lambda_{ijk} L_i L_j e^c_k + \lambda_{ijk}' Q_i L_j d^c_k,
\end{equation}
where $\lambda_{ijk}$ is antisymmetric under the interchange of $i$ and $j$.\footnote{Note that each $L_{i}$ is an $SU(2)_{L}$ doublet. Hence, $L_{i}L_{j}=\epsilon_{AB}L^{A}_{i}L^{B}_{j}$ is antisymmetric in $ij$.} This is accomplished by considering $H_d$ as a fourth generation lepton. In this case, the $\mu$- and $\epsilon_i$-terms can be combined to read $\mu_m \hat L_m' H_u$, where $m=0,\dots,3$, $\hat L_0' = H_d$, $\hat L_{1,2,3}' = L_i$, $\mu_0 = -\mu$ and $\mu_{1,2,3} = \epsilon_i$. The $\mu_m$ term can be perturbatively rotated so that only $\mu_0$ is nonzero. This requires the rotation $\hat L' \to \hat L = R_\mu \hat L' $ with
\begin{equation}
	R_\mu =
	\begin{pmatrix}
		1 & -\frac{\epsilon_1}{\mu} & -\frac{\epsilon_2}{\mu} & -\frac{\epsilon_3}{\mu}
	\\
		\frac{\epsilon_1}{\mu} & 1 & 0 & 0
	\\
		\frac{\epsilon_2}{\mu} & 0 & 1 & 0
	\\
		\frac{\epsilon_3}{\mu} & 0 & 0 & 1
	\end{pmatrix}.
\end{equation}
Implicit in this is that $\epsilon_i \ll \mu$, which follows from the fact that $\epsilon_i$ contributes to neutrino masses, as we shall see later. The rotation leaves only one bilinear between $H_u$ and a linear combination of $L_m'$, which is, of course, mostly composed of $H_d$. This rotation must also be applied to $H_d$ in the down-type quark Yukawa term, $Y_d$, and the charged lepton Yukawa coupling term, $Y_e$, see Eq.~(\ref{W}). The parameterization of $\lambda_{ijk}$ and $\lambda_{ijk}'$ can be read off from this rotation:
\begin{align}
	\label{lambda}
	\lambda_{ijk} & = \frac{1}{2}{Y_e}_{ik}\frac{\epsilon_j}{\mu}-\frac{1}{2}{Y_e}_{jk}\frac{\epsilon_i}{\mu}
	\\
	\lambda_{ijk}' & = {Y_d}_{ik}\frac{\epsilon_j}{\mu}. \label{lambda2}
\end{align}
Because the charged lepton and down quark Yukawa matrices are dominated by the three-three component which gives mass to the tau lepton and bottom quark respectively, those matrices can be calculated to be $Y_e\sim \text{diag}(0,0,Y_\tau)$ and $Y_d \sim \text{diag}(0,0,Y_b)$. This means that the largest elements in the trilinear RPV Yukawas are $\lambda_{3i3} = -\lambda_{i33} = Y_\tau\epsilon_i/\mu $ and $\lambda_{3i3}' = Y_b \epsilon_i/\mu$.

As an application of this rotation, consider the lepton number violating decay $\mu \to e \gamma$. This places the following approximate bound on the trilinear $R$-parity violating couplings~\cite{Barbier:2004ez}:
\begin{equation}
	| \lambda_{23k} \lambda_{13k}| \lesssim 2 \times 10^{-4} \left(\frac{m_{\tilde \nu_3}}{100 \text{ GeV}}\right)^{-2} \ .
\end{equation}
Using Eq.~(\ref{lambda}) yields
\begin{equation}
	\left|\frac{\epsilon_1 \epsilon_2 }{\mu^2}\right| \lesssim 2.5\times 10^{-3} \left(\frac{m_{\tilde \nu_3}}{100 \text{ GeV}}\right)^{-2}
\end{equation}
as the most stringent constraint. This corresponds to  $\tan \beta = 55$, approximately the upper bound on $\tan \beta$ that keeps $Y_\tau$ perturbative up to the GUT scale. The dependence on $\tan \beta$ is due to the fact that the SUSY Yukawa coupling $Y_\tau = \sqrt 2 \, m_\tau/v_d$, where $m_\tau$ is the tau mass. This is negligible due to the suppression of the lepton Yukawa coupling and the $\mu$ term. One would expect $\epsilon_{i}$ values much lower than this bound due to constraints from neutrino masses, as we shall see later. It is worth noting that contributions to $\mu \to e \gamma$ also arise from the $e_i \tilde W^+$ term in Eq.~(\ref{L.n}). However, this is further suppressed due to the $\tilde W^+$-charged lepton mixing, which is proportional to lepton masses. See the approximate value in Eq.~(\ref{eq:wino.mixing}).

Using Eq. (\ref{lambda2}), the decay width of the stop LSP into a bottom quark and a charged lepton (henceforth, referred to as a bottom--charged lepton) is given by
\begin{align}
	\label{x.width}
	\Gamma_{\tilde t_1 \to b \ell^+_i} & \sim \frac{1}{16 \pi}Y_b^2 \left| \frac{\epsilon_i}{\mu} \right|^2 m_{\tilde t_1},
\end{align}
where $\tilde t_1$ indicates the lightest of the two physical stop states (SUSY mass eigenstates are typically numbered from lightest to heaviest). While this neglects order one factors and the contributions from ${v_L}_i$, it is useful for getting an impression of how the stop lifetime depends on the strength of $R$-parity violation. At any rate, it will be shown later that $\epsilon_i$ is typically larger than ${v_L}_i$. An order of magnitude approximation for the lifetime can be simply attained from the largest $\epsilon_i$ value, denoted $\epsilon_\text{max}$, by
\begin{equation}
	\tau_{\tilde t_1} \sim 1\times 10^{-14} \left( \frac{\epsilon_\text{max}/\mu}{10^{-5}}\right)^{-2} \left(\frac{100}{1+\tan^2 \beta} \right) \left(\frac{500 \text{ GeV}}{m_{\tilde t_1}} \right) \text{seconds}.
\end{equation}
Taking representative values of $\mu, m_{\tilde t_1} = 500$ GeV and $\tan \beta =10$ , the lifetimes can be divided up into the following interesting regimes:
\begin{itemize}
	\item Cosmologically significant ($\epsilon_\text{max} \lesssim 10^{-10}$ GeV): The decays of squarks with lifetimes greater than about 100 seconds would disrupt the predictions of big bang nucleosynthesis, see reference~\cite{Kusakabe:2009jt} for example, and would therefore be ruled out.
	\item Collider stability ($10^{-10} \text{ GeV } \lesssim \epsilon_\text{max} \lesssim 10^{-7} \text{ GeV }$): In this regime, the decay length of the squark is longer than the radius of the LHC detectors, about ten meters in size. Such squarks would hadronize and are referred to as $R$-hadrons. These states would be detectable through their activity in the hadronic calorimeter of the detectors and have been studied in references~\cite{Raby:1997pb, Berger:2003kc,Buckley:2010fj,Aad:2011yf,Aad:2012zn,Aad:2012pra}, for example.
	\item Displaced vertices ($10^{-7} \text{ GeV } \lesssim \epsilon_\text{max} \lesssim 10^{-4} \text{ GeV}$): Squark decays inside an LHC detector with a decay length greater than a millimeter have a large enough displaced vertex from the squark origin to be measured. Such vertices, in a phenomenologically similar scenario, were discussed in~\cite{Graham:2012th}. Experimentally, some searches for displaced vertices have been performed in references~\cite{Aad:2011zb,Aad:2012zx,Chatrchyan:2012jwg}.
	\item Prompt decays ($\epsilon_\text{max} \gtrsim 10^{-4} \text{ GeV}$): Decays in this case occur at an indistinguishable distance from the collision point at an LHC detector.
\end{itemize}

The physics associated with non-prompt decays is mostly dependent on the mass of the squark (through its production) and its decay length (displaced vertices or collider stable squarks). Such probes would not be the ideal way of studying the specific branching ratios of the squarks predicted in the model under consideration. In addition such signals have already been analyzed in the references above. We therefore continue this paper considering prompt squark LSP decays only. As we shall see, this will intimately relate the neutrino sector to $R$-parity violation.

The existence of this relationship is already suggested by Eqs.~(\ref{W.brpv}) and (\ref{L.n}). These RPV bilinear terms mix fields with different $R$-parity number but the same spin and SM quantum numbers. Specifically, the neutrinos now mix with the neutralinos, Eq.~(\ref{neutralino}), the charged leptons mix with the charginos, Eq.~(\ref{chargino}) and the Higgs fields mix with the sleptons. The neutrino/neutralino mixings are crucial because they generate tree-level Majorana neutrino masses through a seesaw mechanism. As a result of this, the bilinear $R$-parity violating terms cannot be too large. All $R$-parity violating effects will therefore be negligible compared to the $R$-parity conserving effects, except for the LSP, which now decays via RPV. 

Since $R$-parity violation simultaneously determines both the neutrino sector and the decays of the LSP, it is possible that some of the information from the neutrino sector will be revealed in the LSP decay. This is an exciting and rare opportunity to relate these two fields.

\section{Neutrino Masses and R-parity Violation}
\label{sec.nu.mass}
Any model with right-handed neutrinos allows for Dirac neutrino masses through the Yukawa coupling between left- and right-handed neutrinos. In this model, Majorana masses are also possible due to the VEV of the right-handed sneutrino. As mentioned above, only one generation of right-handed sneutrino can attain a significant VEV~\cite{Mohapatra:1986aw, Ghosh:2010hy, Barger:2010iv}. This means that lepton number is only significantly violated (TeV-scale violation) in one generation of the right-handed neutrinos. It is only that generation of right-handed neutrinos that will attain a TeV-scale mass. This gives rise to a system of neutrinos with three layers: a TeV scale right-handed Majorana neutrino, the three active neutrinos and two light sterile neutrinos\footnote
{
	Sterile neutrinos are typically sub-MeV fermions without SM quantum numbers. In this model, their masses must be at or below those of the left-handed, or active, neutrinos since their masses arise from Dirac Yukawa couplings to the left-handed neutrinos. Models with two sterile neutrinos are sometimes called 3+2 models in the literature, where the three represents the active neutrinos.
}.

Majorana masses for the active neutrinos are generated through an effective type I seesaw mechanism~\cite{Minkowski:1977sc, Yanagida, GellMann:1980vs, Mohapatra:1979ia} where the seesaw fields include the one heavy right-handed neutrino and the neutralinos. Once the heavy seesaw fields are integrated out, the Majorana contribution to the neutrino mass matrix is
\begin{equation}
	\label{nu.mass.1}
	{m_\nu}_{ij} = A {v_L}_i^* {v_L}_j^* + B \left({v_L}_i^* \epsilon_j + \epsilon_i {v_L}_j^* \right) + C \epsilon_i \epsilon_j \ .
\end{equation}
The non-flavored parameters, $A$, $B$ and $C$, are the results of integrating out the heavy fields. They, and more details, are given in Appendix A. The Dirac neutrino mass contributions are simply given by the product of the up-type Higgs VEV and the neutrino Yukawa couplings that do not couple to the third generation right-handed neutrino: $\frac{1}{\sqrt 2}{Y_\nu}_{i, j \neq 3} v_u$.

One of the main tools at our disposal for probing the neutrino sector is the observation of neutrino oscillations. Such oscillations between two neutrinos are determined by the amount of mixing between the two neutrinos and their mass difference. In a purely Dirac neutrino case, the active-sterile mixing is maximal but the mass difference is zero and, therefore, no active-sterile oscillations result. Here, in the pure Majorana case, the mass difference is significant but the mixing is negligible. A situation in which both Dirac and Majorana mass contributions are comparable would lead to large active-sterile oscillations which have not been observed and are therefore ruled out, \textit{e.g.}~\cite{deGouvea:2009fp, Antonello:2012pq}. 

The question then remains, should this analysis assume that neutrinos receive their masses dominantly from Dirac or Majorana mass terms? Here, already, the connection to $R$-parity becomes important. Prompt LSP decays, which were argued to be of interest in the last section, will allow significant Majorana masses. Since these cannot coexist with significant Dirac masses, neutrinos must receive their masses dominantly from Majorana mass terms. This makes further study of the Majorana mass matrix, Eq.~(\ref{nu.mass.1}), fruitful.

As a first step, it is important to notice that the determinant of the neutrino mass matrix in Eq.~(\ref{nu.mass.1}) is zero. This is a consequence of the flavor structure and is independent of the $A, B$ and $C$ parameters. Closer observation reveals that only one eigenstate is massless. This constrains the neutrino masses to be either in the normal hierarchy (NH):
\begin{equation}
	m_1 = 0 < m_2 \sim 8.7 \text{ meV} < m_3 \sim 50 \text{ meV}
\end{equation}
or in the inverted hierarchy (IH):
\begin{equation}
	m_1 \sim  m_2 \sim 50 \text{ meV} > m_3 = 0
\end{equation}
where only the squared mass differences are measured in neutrino oscillation experiments.

The relevant seesaw contributions from $A, B$ and $C$ are also informative. For example, the term proportional to $A$ in Eq.~(\ref{nu.mass.1}) is a contribution associated with the VEVs of the left-handed sneutrinos. It arises from neutrino-gaugino mixing such as in Eq.~(\ref{L.n}). The gauginos are naturally Majorana due to their soft masses and, therefore, integrating them out directly leads to Majorana mass terms for the neutrinos. One can therefore conclude that
\begin{equation}
	\label{eq:AX}
	A\sim \frac{1}{m_\text{soft}},
\end{equation}
where $m_\text{soft}$ is some combination of gaugino and Higgsino masses. This conclusion can be verified with the full analytic expression for $A$ in Appendix~\ref{nu}. The parameter $C$, on the other hand, arises through neutrino-Higgsino mixing because of the $\epsilon_{i}$ term. Higgsinos are not Majorana particles before electroweak symmetry breaking and only their electroweak mixings with the gauginos gives them a Majorana nature. Therefore, $C$ must include at least two factors of Higgsino-gaugino mixing terms, each of which is proportional to the ratio of an electroweak VEV to $m_\text{soft}$:
\begin{equation}
	\label{eq:CX}
	C \sim \frac{v^2}{m_\text{soft}^3}.
\end{equation}
A similar argument yields that $B\sim v/m_\text{soft}^2$ at lowest order. All of these conclusions can be verified with the full expressions in Appendix~\ref{nu}.

The neutrino mass matrix is diagonalized by the so-called PMNS matrix:
\begin{equation}
	V_{PMNS} = 
	\begin{pmatrix}
		c_{12} c_{13}
		&
		s_{12} c_{13}
		&
		s_{13} e^{-i \delta}
		\\
		-s_{12} c_{23} - c_{12} s_{23} s_{13} e^{i \delta}
		&
		c_{12} c_{23} - s_{12}  s_{23} s_{13} e^{i \delta}
		&
		c_{13} s_{23}
		\\
		s_{12} s_{23} - c_{12} c_{23} s_{13}e^{i \delta}
		&
		-c_{12} s_{23} - s_{12} c_{23} s_{13} e^{i \delta}
		&
		c_{13} c_{23}
	\end{pmatrix} \times \text{diag}(1, e^{i \alpha/2}, 1),
\label{b1}
\end{equation}
where $c_{ab} (s_{ab}) = \cos \theta_{ab} (\sin \theta_{ab})$. There are $N-1$ Majorana phases associated with N Majorana neutrinos. This translates into only one Majorana phase, $\alpha$, in this case because one of the neutrinos is massless and, therefore, does not have a Majorana mass. The CP phase $\delta$ corresponds to the freedom in the three-by-three $Y_\nu$ matrix. In models that predict a massless neutrino, such as the one discussed here, the neutrino masses in terms of the mass squared differences in the {\it normal hierarchy} are
\begin{equation}
	m_1 = 0, \quad m_2 = \sqrt{\Delta m_{21}^2},\quad m_3 = \sqrt{\Delta m_{31}^2},
\label{b2}
\end{equation}
while in the {\it inverted hierarchy} one has
\begin{equation}
	m_1 = \sqrt{\Delta m_{31}^2}, \quad m_2 = \sqrt{ \Delta m_{31}^2+\Delta m_{21}^2 },\quad m_3 = 0.
\label{b3}
\end{equation}
The current values for the parameters in \eqref{b1} and \eqref{b2}, \eqref{b3} are given in~\cite{Tortola:2012te, GonzalezGarcia:2012sz, Fogli:2012ua}. We use the most recent values~\cite{nufit} from the collaboration of reference~\cite{GonzalezGarcia:2012sz}, which at one sigma are given by
\begin{align}
\begin{split}
\label{nu.data}
	\sin^2 &\theta_{12} =  0.306^{+0.012}_{-0.012}, \quad 
	\sin^2 \theta_{23} = 0.446^{+0.007}_{-0.007} \text{ \ or \ } 0.587^{+0.032}_{-0.037}, \quad
	\sin^2 \theta_{13} = 0.0229^{+0.0020}_{-0.0019}, 
\\
	& \Delta m_{21}^2 (10^{-5} \text{ eV}^2)= 7.45^{+0.19}_{-0.16}, \quad
	\Delta m_{31}^2 (10^{-3} \text{ eV}^2)= 2.417^{+0.013}_{-0.013}, \quad
	\delta (^\circ) = 265^{+56}_{-61}.
\end{split}
\end{align}

\vspace{0.2cm}
\noindent Note that at three sigma, $\delta$ spans its full range of $0^\circ - 360^\circ$ and that $\alpha$ has not been measured. The two values of $\theta_{23}$ represent a degeneracy in the best fit to the data.

One can solve for the flavorful parameters $\epsilon_i$ and ${v_L}_i$ by requiring that the diagonalization of the neutrino mass matrix, Eq.~(\ref{nu.mass.1}), yields the correct neutrino data specified in Eq.~(\ref{nu.data}). A procedure for this is outlined in Appendix~\ref{nu} in terms of a new set of variables $E_i$ and $V_i$, where
\begin{align}
	\label{vL}
	{v_L}_i & =  {V_{PMNS}}_{il} \, V_l^*,
	\\
	\epsilon_i & =  {V_{PMNS}^*}_{il} \, E_l.
\end{align}
These imply that $\epsilon_i$ and ${v_L}_i$ should be on the order of magnitude of $E_\text{max}$ and $V_\text{max}$ respectively--where $E_\text{max}$ and $V_\text{max}$ are the largest of $E_i$ and $V_i$--since the elements of $V_{PMNS}$ are mostly of order one. In the normal hierarchy $E_1, V_1 = 0$ and Eqs.~(\ref{eq:826}), (\ref{eq:V2}), and (\ref{eq:V3}) are used to calculate $E_2$ and $V_{2,3}$ in terms of $E_3$. Together, they imply that $V_\text{max}\sim(\mathcal O(1)\frac{B}{A} + \mathcal O(1)\sqrt\frac{C}{A})E_\text{max}$, where the coefficients are of order one as long as there are not finely tuned numerical cancellations between terms. The same conclusion holds in the inverted hierarchy. This in turn means that $v_{L_i}\sim(\mathcal O(1)\frac{B}{A} + \mathcal O(1)\sqrt\frac{C}{A})\epsilon_i$. Based on the approximations made above for $A$, $B$ and $C$ in Eqs.~(\ref{eq:AX}) and~(\ref{eq:CX}), it follows that
\begin{equation}
    |\epsilon_i| \sim \frac{m_\text{soft}}{v} |{v_L}_i| \ .
\end{equation}
Quantitatively $\epsilon_i > {v_L}_i$ is verified through the scan specified in Table~\ref{scan}, which is used to generate the numerical results in the next section. Indeed, we find that for $80 \%$ of the points $\epsilon_i > {v_L}_i$ for all $i$ and that the largest $\epsilon_{i}$ value is larger than the largest $v_{Li}$ value ($\epsilon_\text{max} > {v_L}_\text{max}$) in $97 \%$ of the points. Points that do not satisfy these conditions correspond to finely tuned cancellations between terms which, although unlikely, nevertheless arise randomly in the scan. This indicates that $\epsilon_\text{max}$ typically approximates the amount of $R$-parity violation and that $|\epsilon_i|^2 \gg |{v_L}_i|^2$ is a good approximation. This will be useful to obtain an analytic understanding of the numerical results.

\section{Third Generation Squark LSP's}
\label{stop}

The previous two sections have reviewed various aspects of the minimal SUSY $B-L$ model, RPV and the neutrino sector. It was shown that there is an interesting region of parameter space where the 1) strength of RPV corresponds to prompt LSP decays and 2) where the LSP decays might reveal information about the neutrino sector. This paper plans to study these properties under the assumption that the LSP is a third generation squark; that is, for both a stop and sbottom LSP. In addition, we will place lower bounds on the masses of these sparticles using current publicly available LHC results.

Squark LSPs are interesting in RPV for various reasons. First, they are not possible in RPC, so this provides an opportunity to look beyond the typical SUSY LSP candidates and beyond the typical SUSY signatures. Specifically, squark LSPs behave like leptoquarks, meaning they are scalar particles that are pair produced and decay into a quark and a lepton. The stops and sbottoms have the following possible decays:
\begin{align}
	\tilde t_1 \to t \ \nu_i, \  \  \text{or} \  \  \tilde t_1 \to b \ \ell^+_i \ , \label{PD1}
\\
	\tilde b_1 \to b \ \nu_i, \  \  \text{or} \  \  \tilde b_1 \to t \ \ell_i^- \ , \label{PD2}
\end{align}
where $\tilde t_1$ and $\tilde b_1$ are the lightest physical stop and sbottom respectively.
 
Colored particles are, furthermore, more abundantly produced at the LHC, so more aggressive bounds can be placed on them. Generally, one expects a third generation squark to be lighter than the first two generations on the basis of the renormalization group equations. However, this only holds true if one starts with fairly degenerate squarks in all three generations at some high scale associated with soft SUSY breaking. From a phenomenological point of view, the first two generation of squarks should be relatively degenerate to avoid large disallowed contributions to flavor physics processes. This is known as the SUSY flavor problem. Light degenerate first and second generation squarks effectively double the expected number of events for a given process and will consequently have stronger bounds. Furthermore, the first two generations have additional contributions to their production cross section due to the presence of light quarks in the proton. This can, once again, increase the number of events. For these reasons, we continue our analysis focusing on the third generation squarks. Some general comments about the branching ratios of the first two generations will be made in the discussion.

Stop LSPs are especially compelling because of the central role they play in SUSY. Before discussing this further, we briefly review some basic stop phenomenology. More details can be found in Appendix~\ref{sfermions}. In the gauge eigenstate basis, the stop sector contains the $\tilde t$ field, which is the superpartner of the left-handed top and part of the squark $SU(2)_L$ doublet $\tilde Q$. Since it is a scalar, the stop has no actual chiral properties. The stop sector also contains the superpartner of the right-handed top, $\tilde t^c$, which is an $SU(2)_L$ singlet. Both have unrelated soft squared masses and are mixed through mass mixing terms. Diagonalization yields the physical stops $\tilde t_1$ and $\tilde t_2$, which are traditionally labeled so that $m_{\tilde t_1} < m_{\tilde t_2}$. The mass mixing term leads to what is usually referred to as the left-right mixing angle in the stop sector, $\theta_t$, with the convention used here that $\theta_t = 0^\circ$ ($\theta_t = 90^\circ$) corresponds to a purely left-handed (right-handed) lightest stop, $\tilde t_1$. A purely left-handed $\tilde t_1$ cannot be the LSP because its $SU(2)_L$ partner, the left-handed sbottom, will always be lighter. This is because they share the same SUSY-breaking soft mass squared term and both get $F$-term contributions from their SM partner mass squared. That is, the sbottom mass gets a bottom mass squared contribution and the stop gets a top mass squared contribution. Since the top is much heavier than the bottom, the left-handed stop will always be heavier than the left-handed sbottom.

The stops in SUSY are important because they couple most strongly to the Higgs. This means they contribute most to the little hierarchy problem and provide a measure of the fine-tuning required in SUSY models. In RPC, stop decays can involve complicated decay chains with 
multi-particle final states making determination of the stop mass from the observation of such a decay difficult. As an LSP with $R$-parity violation, stop decays are very clean in the sense that each stop decays to only two particles. Therefore, such decays can be used to deduce the stop mass in a relatively straightforward way. This is especially true for the bottom--charged lepton channel, whose final states are both detectable. Neutrinos, on the other hand, escape the detector as missing energy. As we shall see, typically the bottom--charged lepton channel dominates the stop decays.

The issue of the little hierarchy problem is also strongly linked to the Higgs mass. In SUSY, the Higgs tree-level mass must be less than the $Z$ mass. This can be increased at the loop level by radiative corrections to the Higgs mass which grow as the logarithms of the stop masses and also increase with stop mixing angle. This leads to a conflict between the heavy stops masses needed to make SUSY compatible with the recent Higgs discovery and the desire to keep the stops light so as to minimize fine-tuning in SUSY. The former seems to be an argument against a stop LSP. However, it is possible that only one stop is quite heavy while the second remains light--which will  indeed be the case when the stop mixing angle is relatively large.
This translates into an LSP stop that is composed of significant left- and right-handed components. Since the Higgs mass is not altered in this model, one can consult the MSSM literature to explore the possibilities, \textit{e.g.}~\cite{Carena:2011aa}.

The stop partial widths into top neutrino and bottom--charged lepton are
\begin{eqnarray}
\label{stop.width1}
\Gamma(\tilde t_1 \to t \, \nu_i)&=&\frac{1}{16\pi}(|G^L_{\tilde t_1 t\chi^0_{6+i}}|^2+|G^R_{\tilde t_1 t\chi^0_{6+i}}|^2) m_{\tilde t_1}
	\left(1 - \frac{m_t^2}{m_{\tilde t_1}^2} \right)
	\sqrt{1-2\frac{m_t^2}{m_{\tilde t_1}^2}+\frac{m_t^4}{m_{\tilde t_1}^4}}
	\\
\label{stop.width2}
	\Gamma(\tilde t_1 \to b \, \ell^+_i)&=&\frac{1}{16\pi}(|G^L_{\tilde t_1 b\chi^\pm_{2+i}}|^2+|G^R_{\tilde t_1 b\chi^\pm
_{2+i}}|^2)m_{\tilde t_1},
\end{eqnarray}
where the $G$ parameters are the coefficients of the relevant vertices, $\chi_{6+i}^0 = \nu_i$ and $\chi^\pm_{2+i} = \ell_i^\pm$. They, as well as more details, can be found in Appendix~\ref{FR}. Parametrically, the $G^{L,R}_{{\tilde t}_{1} t\chi^0_{6+i}}$ parameters contain the elements of the matrix that diagonalize the neutrino-neutralino sector and the $G^{L,R}_{{\tilde t}_{1} b\chi^\pm_{2+i}}$ parameters contain the elements of the matrix that diagonalize the lepton-chargino sector and are, therefore, proportional to some combination of $\epsilon_i$ and ${v_L}_i$. Also encoded in the $G$ parameters is information about the stop left-right mixing angle, $\theta_t$.

Before tackling a numerical study of stop LSP phenomenology, it is instructive to approximate the relative sizes of the different branching ratios. This can be done by perturbatively diagonalizing the neutrino-neutralino and charged lepton-chargino mass matrices, as is done in Appendices~\ref{nu} and~\ref{charginos} and applied in Appenedix~\ref{FR}. For ease of comparison, the leading squared amplitudes for the different final states are given in the approximation that $M_{Z_R}^2 \gg m_\text{soft}^2 \gg v^2$. This is a phenomenologically relevant approximation because bounds on $Z_R$ are much higher than electroweak gaugino and Higgsino bounds and both are above the electroweak scale itself. We also employ the results of the last section, $\epsilon_i^2 \gg {v_L}_i^2$. The leading contributions to the square of the vertex amplitude, $|\mathcal{A}|^{2}=|G^{L}|^{2}+|G^{R}|^{2}$, are then
\begin{align}
\label{t1bl}
	|\mathcal{A}(\tilde t_1 \to b \, \ell^+_i)|^2 & \sim c_t^2 Y_b^2 \left|\frac{\epsilon_i}{\mu} \right|^2
\\
\label{t1tnu}
	|\mathcal{A}(\tilde t_1 \to t \, \nu_i)|^2 & \sim
	\left[
		\frac{1}{8}c_t^2
		\left(
			\frac{g_2^2}{M_2} - \frac{g_{BL}^2 g_R^2}{3 M_{\tilde Y}}
		\right)^2
		+ \frac{1}{18}s_t^2
		 \frac{g_{BL}^4 g_{R}^4}{M_{\tilde Y}^2}
	\right]
	\left|
		{V_\text{PMSN}}_{ij}
		\left(
			\frac{v_d \, \epsilon_j}{ \mu}
			+ {v_L}_j^*
		\right)
	\right|^2
	,
\end{align}
where $s_t$ ($c_t$) is $\sin \theta_t$ ($\cos \theta_t$), $M_{\tilde Y} \equiv g_R^2 M_{BL} + g_{BL}^2 M_R$ and there is an implicit sum over $j$. The top--neutrino channel is suppressed compared to the bottom--charged lepton channel both by helicity suppression to the term proportional to $\epsilon_i$ and suppression by ${v_L}_i$ when the lightest stop is not purely right-handed. When the lightest stop is purely right-handed, the leading order bottom--charged lepton amplitude vanishes and the next order term becomes important:
\begin{align}
\label{tr.dk}
	|\mathcal{A}(\tilde t_1 \to b \, \ell^+_i)|^2 \bigg |_{\theta_t \sim 90^\circ} & \sim Y_t^2 \left|\frac{m_{\ell_i} \, {v_L}_i}{\mu \, v_d} \right|^2.
\end{align}
This term is suppressed by both ${v_L}_i$ and the mass of the charged lepton in the final state, ${m_\ell}_i$, indicating that, for the mostly right-handed stop, only the top--neutrino and bottom-tau channels are significant. The stop branching ratios, where branching ratio is defined as the partial width normalized to the total width, falls into two regimes of interest depending on the composition of the stop:
\begin{itemize}
	\item Admixture stop LSP: Stop decays into into bottom--charged leptons dominate, $\sum_i \Gamma(\tilde t_1\to b \ell^+_i) \gg \sum_i \Gamma(\tilde t_1\to t \nu_i) $. We therefore approximated the total width as coming completely from the charged leptons, and the decays of the stop can be described by three branching ratios, which must satisfy
	\begin{equation}
		\text{Br}(\tilde t_1 \to b \, e^+) + \text{Br}(\tilde t_1 \to b \, \mu^+) + \text{Br}(\tilde t_1 \to b \, \tau^+)=1.
\label{admixturestop}
	\end{equation}
	\item Right-handed stop LSP: Only the top--neutrino and bottom-tau channel are significant. We therefore approximate the width as coming completely from these two channels and the decays can be described by two branching ratios, which must satisfy:
	\begin{equation}
		\text{Br}(\tilde t_1 \to b \, \tau^+) + \text{Br}(\tilde t_1 \to t \, \nu)=1.
	\end{equation}
\end{itemize}

Let us qualitatively understand these results, which may be a bit counterintuitive. Since $\epsilon_i$ mixes $\tilde H_u$ with $L_i$, one would expect the leading contributions to be proportional to the $Y_t$, since it couples the stops to $\tilde H_u$ and through it to the $\epsilon_{i}$ parameter. However, such decays are helicity suppressed by a factor of $v^2/m_\text{soft}^2$ (in Eq.~(\ref{t1tnu})) and are, therefore, subdominant. The dominant channel to RPV then usually goes through $\tilde H_d$ and, therefore, includes a factor of $Y_b \epsilon_{i}$. This explains Eq.~(\ref{t1bl}). The top--neutrino channel cannot, however, be accessed through $\tilde H_d$ and must, therefore, suffer the helicity suppression or be suppressed by ${v_L}_i$, as are the two terms in Eq.~(\ref{t1tnu}). The right-handed stop also cannot access $\tilde H_d$. Its decay into bottom--charged lepton must go through $\tilde H_u-\tilde H_d$ mixing and finally through $Y_{e_i} {v_L}_i \tilde H_d^- e_{i}^c$, which is the reason that Eq.~(\ref{tr.dk}) depends on the lepton mass.

With these guidelines in mind, we proceed to our numerical study.
\subsection{Stop LSP Decays and the Neutrino Spectrum}
\label{dks.nuspec}
The numerical procedure starts with the process in Appendix \ref{nu}, which takes as input the unmeasured CP violating phases of the neutrino sector, the neutralino spectrum, the $B-L$ parameters, any one of the $\epsilon_i$ parameters, and two signs. It yields values for ${v_L}_i$ and the other two $\epsilon_i$ that are consistent with neutrino physics. These values are then used to numerically diagonalize the neutrino/neutralino and charged lepton/chargino mass matrices. These rotation matrices are then inputted into the Feynman rules in Appendix~\ref{FR}, which can be used in Eqs.~(\ref{stop.width1}) and (\ref{stop.width2}) to calculate the partial widths. Because of the dependence on a variety of parameters, full analytic relationships between the input parameters and the stop decay branching ratios are complicated and not very illuminating. However, random scans in the space of the input parameters yield fairly simple behavior.

The parameters of our scan and their ranges are specified in Table~\ref{scan}. As mentioned above, the neutrino sector specifies all but one $R$-parity violating parameter, which we choose to be $\epsilon_i$ and we randomly choose the generation, $i$, of $\epsilon_i$ to avoid any bias in the scan. The sign factors, $\zeta_0$ and $\zeta_3$ are further discussed in Appendix \ref{nu}. While only the gluino mass range is shown, we use the GUT inspired gaugino mass relation $M_R:M_{BL}:M_2:M_3 \sim 1:1:2:5$ for the gaugino masses~\cite{Ovrut:2012wg}. This is based on the ratio of the gauge couplings at the TeV scale. The lower ranges on $M_3, \, M_{Z_R}, \, \mu$ and $m_{\tilde t_1}$ roughly correspond to the lower bounds on those particles, while $\mu$ roughly corresponds to the mass of one of the physical chargino states. The lower and upper bounds on $\tan \beta$ are based on keeping all Yukawa couplings perturbative to the GUT scale. Meanwhile, the bounds on $\epsilon_i$ follow from requiring no fine-tuning in the neutrino sector, the conditions for which are described in Appendix~\ref{nu}. This fine-tuning depends on the actual parameter point and we find that non fine-tuned points lie in the range $10^{-4} \text{ GeV}< |\epsilon_i| < 1 \text{ GeV}$, which is used in the scan.

In addition, the uncertainties on the neutrino parameters themselves can quantitatively alter the results. We, therefore, also scan over the three sigma range of the neutrino parameters based on their values and uncertainties given in Eq.~(\ref{nu.data}). To do this, we need a probability distribution to describe the uncertainty in these parameters. A simple Gaussian will not do, because the uncertainties in some of the neutrino parameters are asymmetric. Instead we randomly select, with probability one half, which side of the central value a parameter will be on. Then a value for that parameter is randomly generated based on a Gaussian distribution whose standard deviation is equal to the 1$\sigma$ uncertainty on the chosen side of that parameter's central value. The Gaussian distribution is curtailed a distance of three standard deviations away from the central value. No correlations between neutrino parameter ranges are taken into account here. Furthermore, the CP-violating phases, $\delta$ and $\alpha$, are scanned over their full range and the central value of $\theta_{23}$ used is randomly chosen between the two ambiguous experimental values.

Since we are studying a stop LSP, points in the scan at which one of the neutralinos or charginos end up being lighter than the stop are rejected. It is also possible that some points in the scan may have a nearly purely left-handed lightest stop, which may be unable to be the LSP (see Appendix~\ref{sfermions}). A criterion for excluding such points from the scan would depend on parameters that do not effect the physics of this paper, so we do not impose it here. Such a criterion would have no impact on the overall trends displayed by our scan, so it would not effect the conclusions of this paper.
\begin{table}[htdp]
\begin{center}
\begin{tabular}{|c|c|}
\hline
\hspace{1cm} Parameter \hspace{1cm} & \hspace{0.7cm} Range \hspace{0.7cm}
\\
\hline
$M_3$ (TeV) 	& 1.5 \, -- \, 10
\\
$M_{Z_R}$ (TeV)	&  2.5 \, -- \, 10
\\
$\tan \beta$ & 2 \, -- \, 55
\\
$\mu$ (GeV)& 150 \, -- \, 1000
\\
$m_{\tilde t_1}$ (GeV)& 400 \, -- \, 1000
\\
$\theta_{t}(\vphantom{t}^\circ)$ & 0 \, -- \, 90
\\
$| \epsilon_i| $ (GeV) & $10^{-4}$ \, -- \, $10^{0}$
\\
$\arg \left(\epsilon_i \right)$ & 0 \, -- \, 360
\\
$i$ & 1 \, -- \, 3
\\
$\zeta_0, \, \zeta_3 $ & -1,\ 1
\\
$\delta, \, \alpha(\vphantom{t}^\circ)$ & 0 \, -- \, 360
\\
Neutrino Hierarchy & NH, IH
\\
\hline
\end{tabular}
\end{center}
\caption{Ranges for the parameter scan. The neutrino sector leaves only one unspecified $R$-parity violating parameter, which is chosen to be $\epsilon_i$ where the generational index, $i$, is also scanned to avoid any biases. The scanned gluino mass is shown here, while the other gaugino masses are extrapolated from the GUT relation $M_R:M_{BL}:M_2:M_3 = 1:1:2:5$.}
\label{scan}
\end{table}

\begin{figure}[h]
	\includegraphics[scale=0.4]{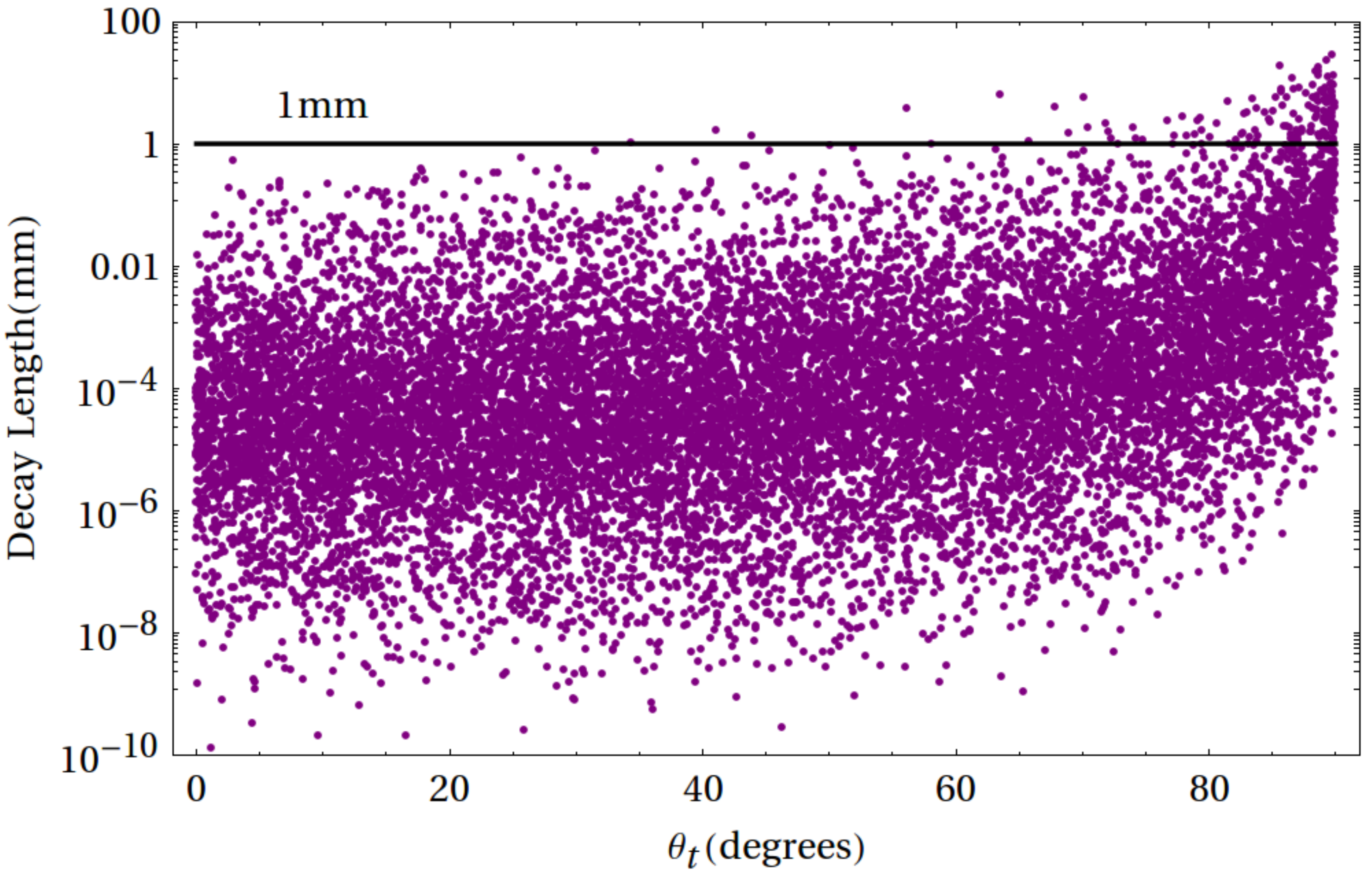}
	\caption{Stop LSP decay length in millimeters versus stop mixing angle. The decay length increases sharply past $80^\circ$, where the stop is dominantly right-handed, due to the suppressed right-handed stop decays, Eq.~(\ref{tr.dk}).}
	\label{fig:dk.length}
\end{figure}

We note that due to the extra suppression in the decays of the right-handed stop, Eq.~(\ref{tr.dk}), the LSP stop lifetime increases by a significant amount when it approaches a purely right-handed stop composition. Using the scan from Table~\ref{scan}, we plot the decay length of the stop LSP versus stop mixing angle in Fig.~\ref{fig:dk.length}. The figure shows that for a pure right-handed stop LSP, a significant number of points in the scan yield lifetimes long enough for displaced vertices (decay length greater than a millimeter). We continue our analysis focusing on prompt decays.

\begin{figure}[h]
\includegraphics[scale=0.4]{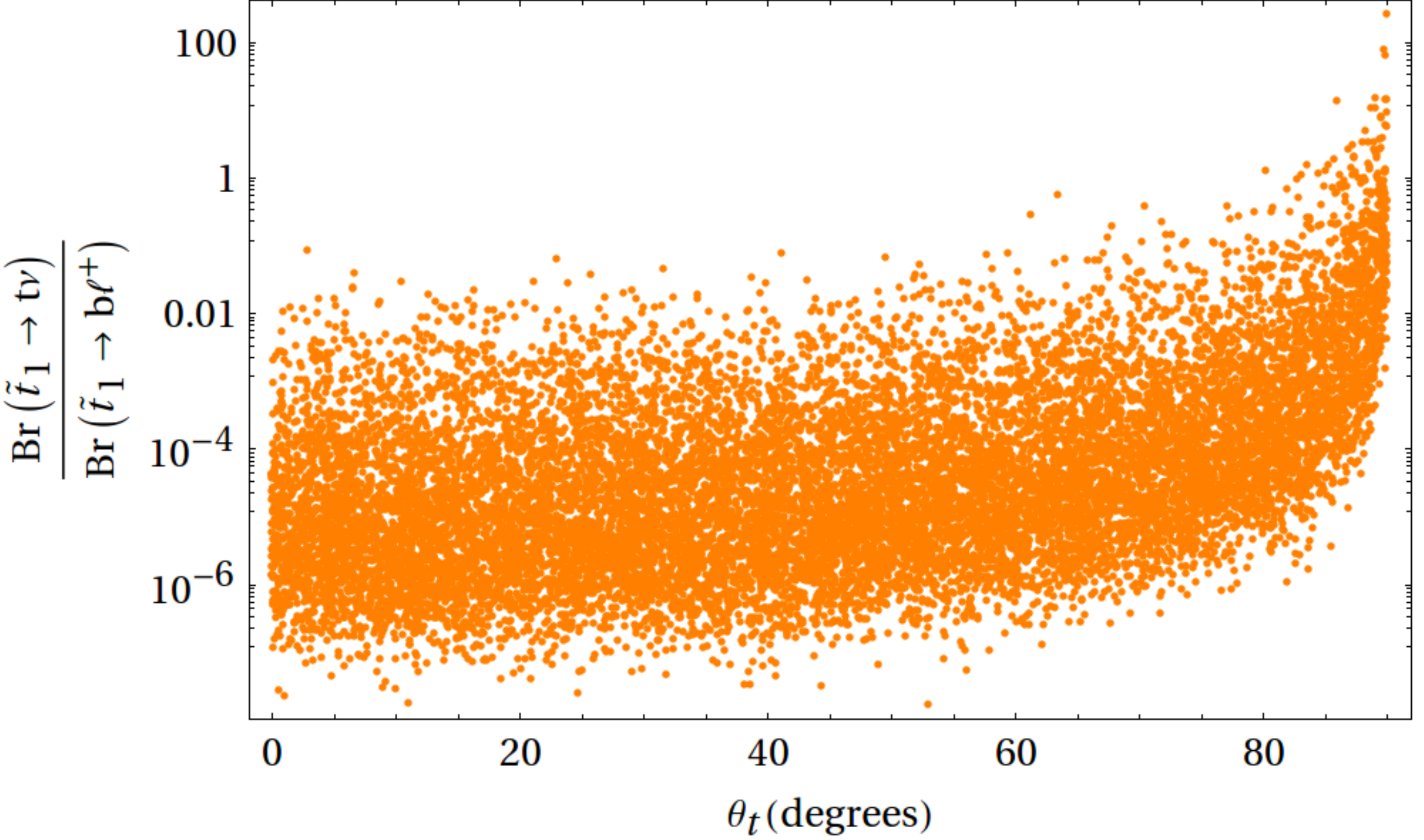}
\caption{$\frac{\text{Br}(\tilde t_1\to t\nu)}{\text{Br}(\tilde t_1 \to b\ell^+)}$ versus stop mixing angle, where $\text{Br}(\tilde t_1\to b\ell^+)\equiv\sum\limits_{i=1}^3\text{Br}(\tilde t_1\to b\ell^+_i)$. For the admixture stop, the branching ratio to $b\ell^+$ is dominant and the branching ratio to $t\nu$ is insignificant for LHC purposes. For a mixing angle greater than about $80^\circ$, corresponding to a mostly right-handed stop, the branching ratio to $t\nu$ can be significant.}
\label{thetat}
\end{figure}

Figure~\ref{thetat} shows how $\text{Br}(\tilde t_1\to t\nu)/\text{Br}(\tilde t_1 \to b\ell^+)$, where $\text{Br}(\tilde t_1\to b\ell^+)\equiv\sum\limits_{i=1}^3\text{Br}(\tilde t_1\to b\ell^+_i)$, depends on the stop mixing angle. This verifies the relationship between the stop mixing angle and branching ratios into bottom--charged lepton and top--neutrino derived from Eqs.~(\ref{t1bl})~-~(\ref{tr.dk}). Figures~\ref{fig:dk.length} and~\ref{thetat} both show that the right-handed stop-like behavior, significant top--neutrino channel and longer lifetimes, turns on around $\theta_t = 80^\circ$.

\begin{figure}[h]
	\includegraphics[scale=0.5]{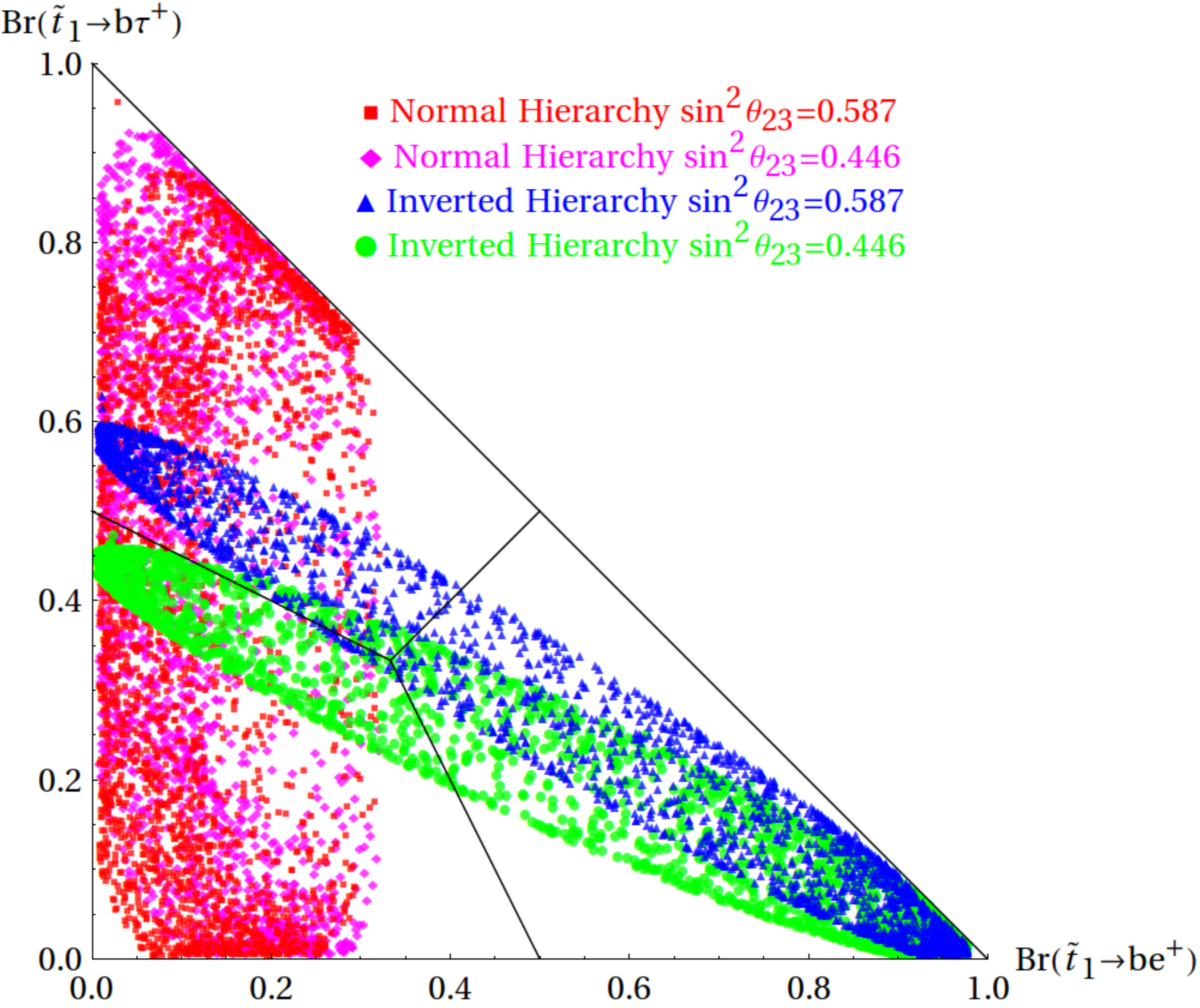}
	\caption{The results of the scan specified in Table~\ref{scan}, but with central values for the measured neutrino parameters in the $\text{Br}(\tilde t_1 \to b \, \tau^+)$ - $\text{Br}(\tilde t_1 \to b \, e^+)$ plane. Due to the relationship between the branching ratios, the $(0,0)$ point on this plot corresponds to $\text{Br}(\tilde t_1 \to b \, \mu^+)=1$. The plot is divided into three quadrangles, each corresponding to an area where one of the branching ratios is larger than the other two. In the top left quadrangle, the bottom--tau branching ratio is the largest; in the bottom left quadrangle the bottom--muon branching ratio is the largest; and in the bottom right quadrangle the bottom--electron branching ratio is the largest. The two different possible values of $\theta_{23}$ are shown in blue and green in the IH (where the difference is most notable) and red and magenta in the NH.}
	\label{fig:Brs.central}
\end{figure}

Perhaps the most striking result from this scan is the connection between the stop decays and the neutrino hierarchy. This connection is evident in Fig.~\ref{fig:Brs.central} where the possible branching ratios are displayed in the $\text{Br}(\tilde t_1 \to b \, \tau^+)$ - $\text{Br}(\tilde t_1 \to b \, e^+)$ plane and where, for simplicity, we start with only the central values of the measured neutrino parameters, Eq.~(\ref{nu.data}). The figure includes only points with $\text{Br}(\tilde t_1\to t\nu)<0.01$. Such points correspond to admixture stop LSP, according to Fig.~\ref{thetat}. Using the top--neutrino branching ratio, instead of the stop mixing angle, to distinguish between the admixture and right-handed stop LSP is preferable because the top--neutrino branching is easier to measure. This means that $\text{Br}(\tilde t_1 \to b \, e^+) + \text{Br}(\tilde t_1 \to b \, \mu^+) + \text{Br}(\tilde t_1 \to b \, \tau^+)=1$ (Eq.~(\ref{admixturestop})), so that the $(0,0)$ point on this plot corresponds to $\text{Br}(\tilde t_1 \to b \, \mu^+)=1$. The reader may observe that Fig.~\ref{fig:Brs.central} includes a small number of points that do not follow the trend displayed by the bulk of the points, and are instead skewed in the direction of larger bottom--tau branching ratio. These rare points correspond to a transitional region between admixture stop and purely right-handed stop where Eq.~(\ref{tr.dk}) is starting to become valid, favoring a larger bottom--tau ratio due to the tau being the heaviest of the lepton. Points that do not satisfy the fine-tuning criteria of the neutrino sector, Eqs.~(\ref{finetuningb}) and~(\ref{finetuninga}), are excluded.

Figure~\ref{fig:Brs.central} is divided into three quadrangles each corresponding to an area where one of the branching ratios is larger than the other two. In the top left quadrangle, the bottom--tau branching ratio is the largest; in the bottom left quadrangle the bottom--muon branching ratio is the largest; and in the bottom right quadrangle the bottom--electron branching ratio is the largest. Recall that the fit to the neutrino data allows two values of $\theta_{23}$. One is shown in blue and and the other in green in the inverted hierarchy (where the impact on stop decays is most notable) and in red and magenta in the normal hierarchy.

Figure~\ref{fig:Brs.central} shows the strong connection between the stop branching ratios and the neutrino sector. The most interesting connection is to the neutrino mass hierarchy. If these decays were observed at the LHC and their branching ratios measured, then it might be possible to determine the neutrino hierarchy, an open question being actively pursued in neutrino physics today~\cite{Cahn:2013taa}.

\begin{figure}[h]
	\includegraphics[scale=0.5]{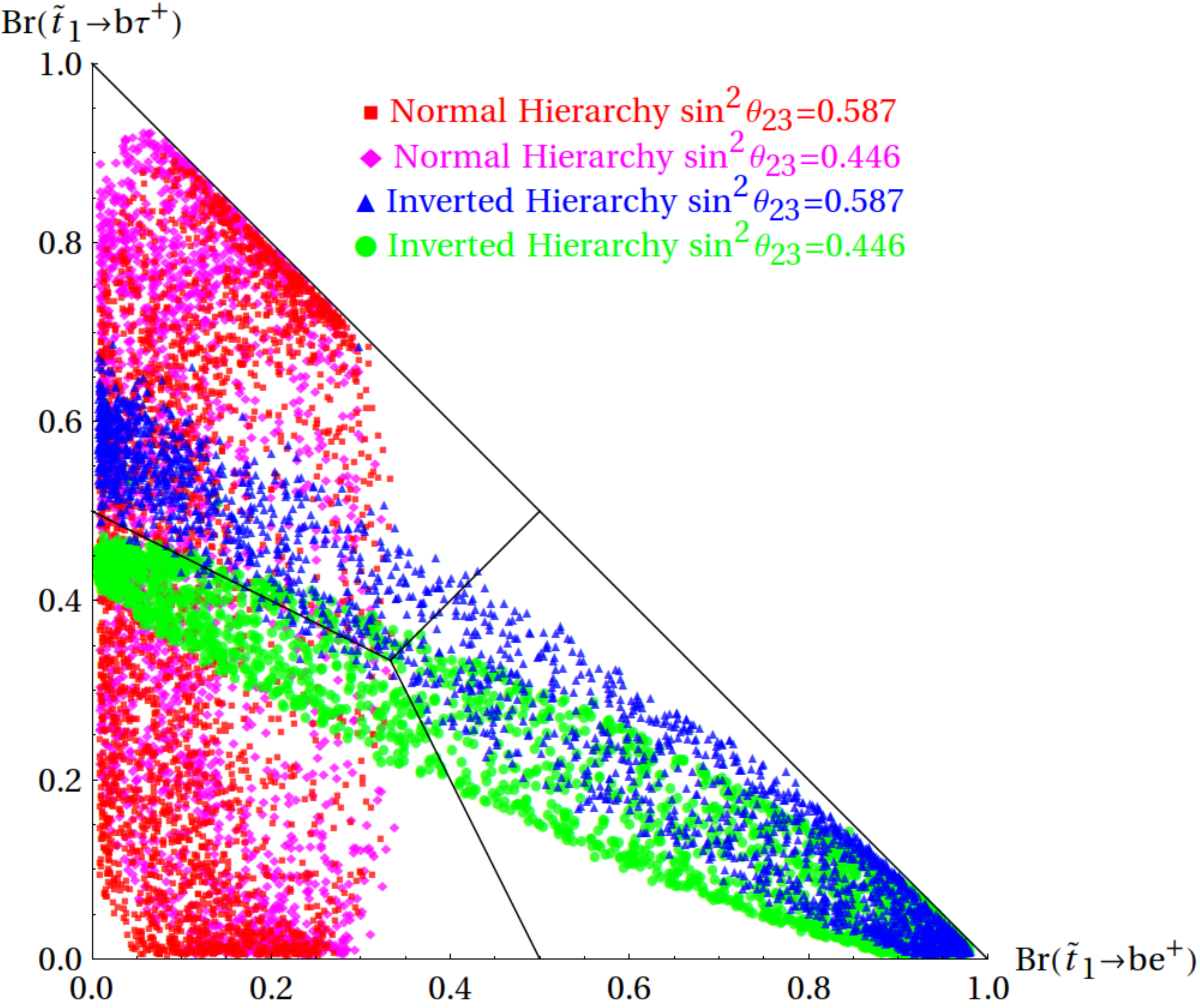}
	\caption{Same as Fig~\ref{fig:Brs.central} except with a Gaussian distributed scan over the neutrino parameters as described in Eq~(\ref{nu.data}).}
	\label{fig:Brs.3sigma}
\end{figure}

The full results including the three sigma scan over neutrino parameters are displayed in Fig.~\ref{fig:Brs.3sigma}. The features of this figure are very similar to those of Fig.~\ref{fig:Brs.central}.  While taking the three sigma range of the neutrino parameters into account has obscured things somewhat compared to Fig.~\ref{fig:Brs.central}, the connection to neutrino physics is still strong and very visual and the conclusions still of interest\footnote{Note that the limited capability of the LHC detectors to precisely measure such branching ratios may also smear out this picture.}. Therefore, assuming one is lucky enough to discover a particle decaying in this way at the LHC, one can then use the measured branching ratios to conclude the following.
\begin{itemize}
	\item If the branching ratio to bottom--electron is the largest branching ratio, the neutrino mass hierarchy is likely to be the inverted hierarchy.
        \item If the branching ratio to bottom--muon is found to be highly dominant, then neutrino masses are likely to be in a the normal hierarchy. If this branching ratio is only slightly dominant, the hierarchy cannot be determined from from this measurement alone, because it is compatible with both normal and inverted hierarchy. However, if the hierarchy were determined to be inverted from some other experiment, this measurement would favor the central value of $\sin^2\theta_{23}\sim0.446$ over $\sin^2\theta_{23}\sim0.587$.
        \item The case where the branching ratio to bottom--tau is highly dominant, the normal hierarchy is favored. If it is only slightly dominant, neither hierarchy is favored, but the central value of $\sin^2\theta_{23}=0.587$ would be slightly favored over $\sin^2\theta_{23}=0.446$ if the hierarchy were determined to be inverted from some other experiment.
	\item A really lucky scenario would land the observer in the electron dominated quadrangle at the top of the blue points or the bottom of the green points. From this, one would be able to argue that the central value of $\sin^2 \theta_{23}$ is closer to 0.587 for the former scenario and $0.446$ for the latter in addition to an inverted hierarchy.
	\item Nature placing us in the white spaces would strongly suggest that this model is not the correct interpretation of the data. One caveat to this is the transition range between an admixture stop LSP and a purely right-handed stop LSP. This might allow some points in the upper white regions but, we found them to be rare in our scan.

\end{itemize}
The above conclusions relate decays that could be observable at the LHC to the neutrino mass hierarchy, which is currently at the forefront of neutrino frontier~\cite{Cahn:2013taa}. Furthermore the hierarchy has important consequences for experiments seeking to measure neutrinoless double beta decay\footnote{
A positive measurement of neutrinoless double beta decay is a clear measurement of lepton number violation and the Majorana nature of neutrinos.
},
which is more prominent in the inverted hierarchy. Measurement of stop LSP decays could allow a prediction of what hierarchy should be found by such experiments. Conversely, if neutrino experiments are able to determine the neutrino mass hierarchy, this could be used to further constrain the types of decays predicted for the LHC.

\begin{figure}[h]
	\includegraphics[scale=0.5]{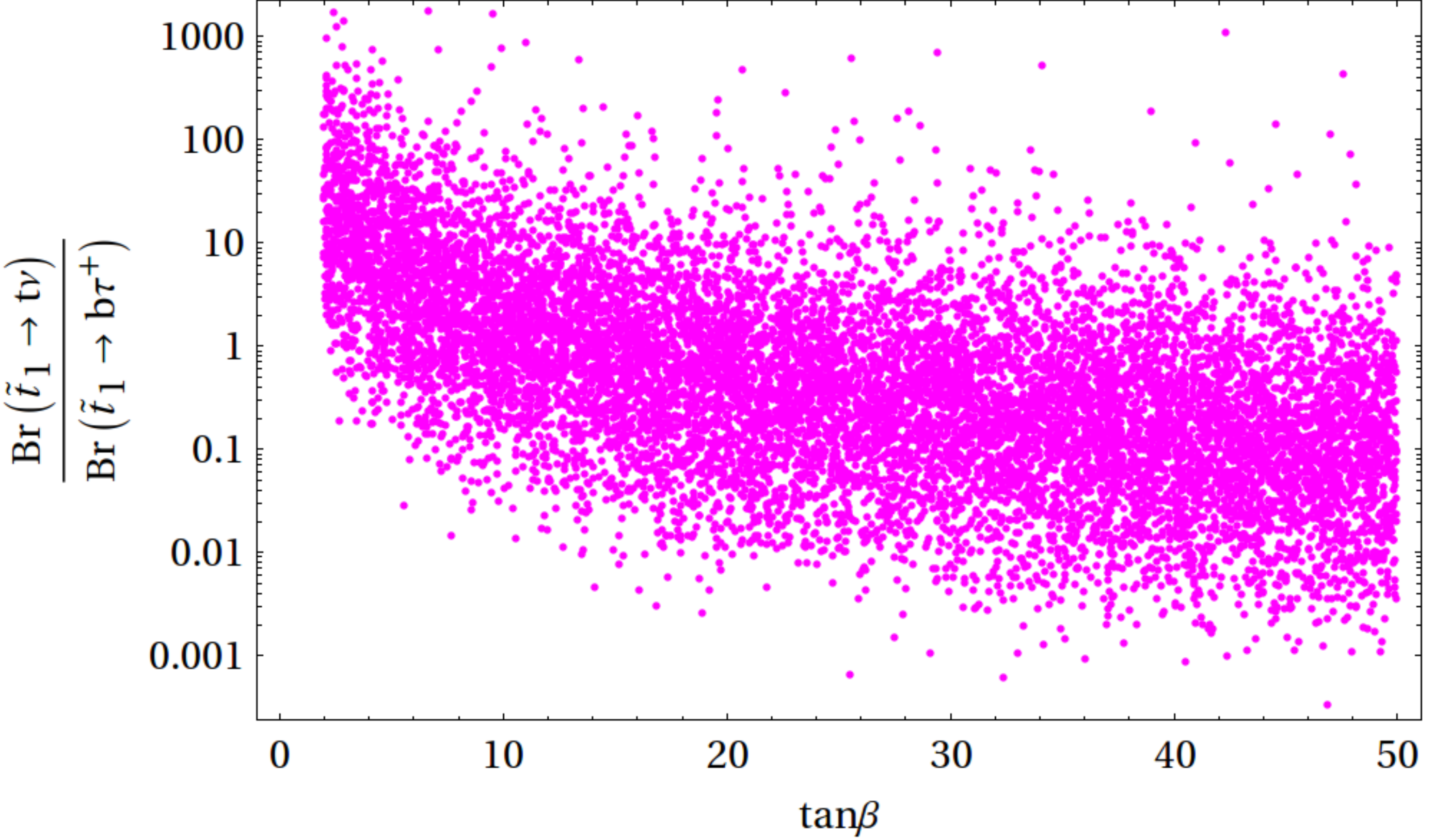}
	\caption{The ratio of the branching ratio of right-handed stops into top--neutrino to the branching ratio of right-handed stops to bottom--tau versus $\tan \beta$. Branching ratios to the lighter charged leptons are suppressed by their masses and therefore negligible in this case. The plot shows a dependence on $\tan \beta$ with small (large) $\tan \beta$ values corresponding to dominant top neutrino (bottom--tau) branching ratio.}
	\label{fig:tr.Br}
\end{figure}
Much past the $\theta_t = 80^\circ$ mark, as seen in Figs.~\ref{fig:dk.length} and~\ref{thetat}, the lightest stop is dominantly right-handed and the connection to neutrino physics is lost. This is because the branching ratios into the lighter generations of leptons are suppressed, Eq.~(\ref{tr.dk}), and because the neutrino generation cannot, of course, be measured at the LHC. Still, in this case, there is an interesting connection between the two decay channels and $\tan \beta$ as can be seen from Eq.~(\ref{tr.dk}). From this, one would expect the bottom--tau channel to dominate at large $\tan \beta$ while the top neutrino channel dominates for low $\tan \beta$. Utilizing the same scan as in Table~\ref{scan} but with $\theta_{\tilde t} = 90^\circ$ produces Fig.~\ref{fig:tr.Br}, which displays $\text{Br}(\tilde t_1 \to t\nu)/\text{Br}(\tilde t_1 \to b \tau^+)$ versus $\tan \beta$. The results confirm the relationship between the branching ratios and $\tan \beta$. 

\subsection{Stop LSP Lower Bounds}
\label{sec:stop.lower.bounds}
 
\begin{figure}[h]
\includegraphics[scale=0.8]{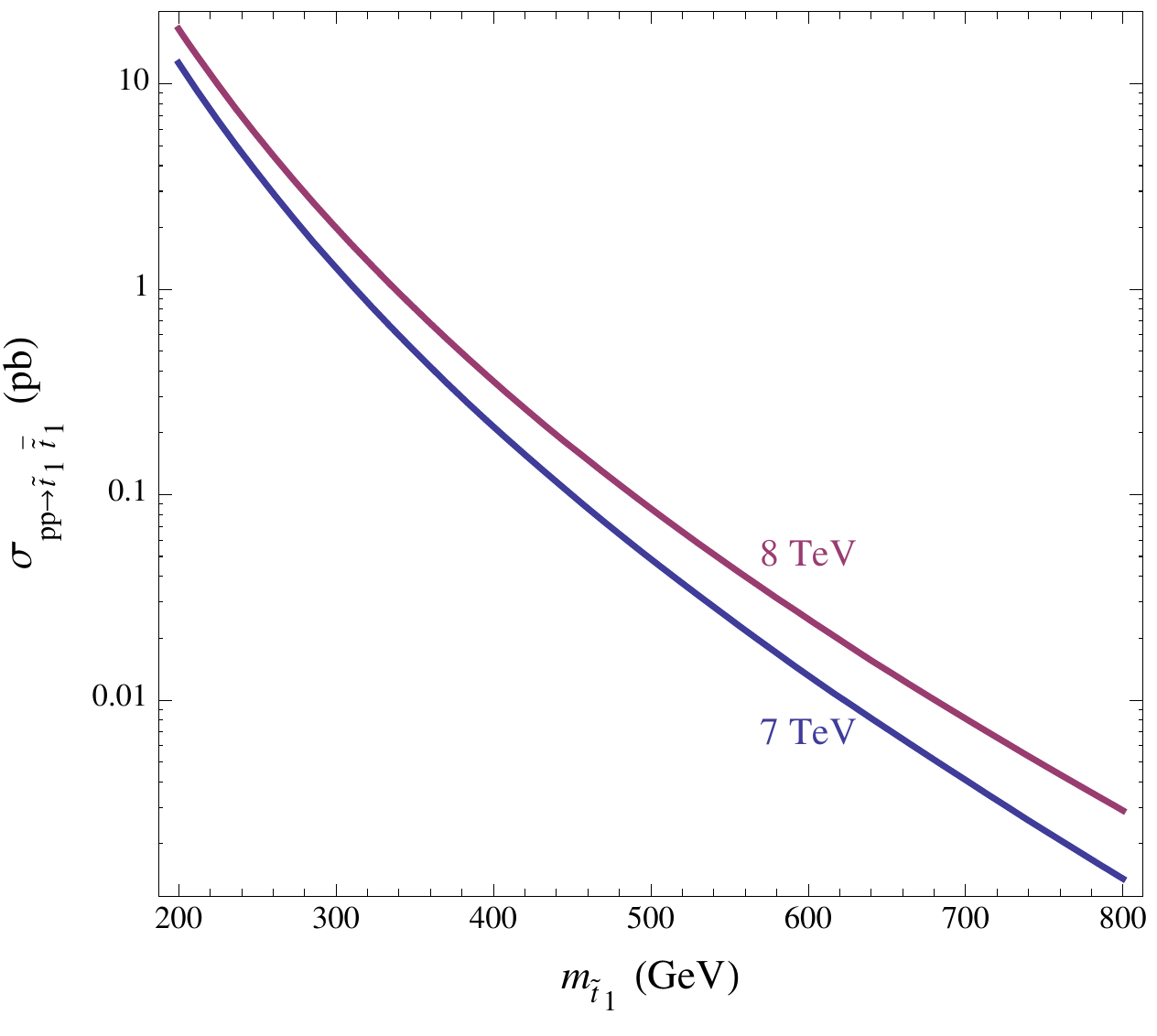}
\caption{Stop pair production cross section at the 7 and 8 TeV LHC as calculated by the ATLAS, CMS and LPCC SUSY working group.}
\label{fig:stop.cross.section}
\end{figure}

LHC searches that place limits on one of the final states discussed previously can be reinterpreted to place lower bounds on the stop mass. Naively, bounds on the stop mass can be placed based on the number of expected events, for a given mass, as compared to the number of observed events. Of course, realistically, one must also take the background for the process into account as well various detector level details. Putting these aside for the moment, the number of expected events depends only on the mass of the stop, its branching ratios and the center of mass energy. Squarks are always pair produced in this model and, in the admixture case, result in the final state $b \, \bar b \ell_i^-  \ell_j^+$. The number of such events is given by
\begin{equation}
	L \times \left(2 - \delta_{ij} \right) \times \sigma_{pp\to \tilde t_1 \bar{\tilde t}_1} \times \text{Br}(\tilde t_1 \to b \ell^+_i) \times \text{Br}(\tilde t_1 \to b \ell^+_j),
\end{equation}
where $L$ is the luminosity (the most recent LHC run has 20$^{-1}$ fb of luminosity) and $\sigma_{pp\to \tilde t_1 \bar{\tilde t}_1}$ is the hadron level cross section, which results from summing partonic contributions. These partonic contributions are a product of the parton level cross section and the appropriate parton distribution function (PDF) integrated over the parton's momentum fraction of the hadron's momentum. For LHC stop production, the leading order parton contributions come from gluon fusion and quark-quark fusion. The parton-level cross section formulas can be found in~\cite{Dawson}. Here we plot the production cross section at next to leading order in $\alpha_S$, including resummation at next-to-leading log, as calculated by the ATLAS, CMS and LPCC SUSY working group~\cite{Kramer:2012bx, Kramer2}, as a function of stop mass at both a 7 and 8 TeV LHC, in Fig.~\ref{fig:stop.cross.section}.

Leptoquarks exists in various extensions of the standard model, such as unification and partial unification models, and have been searched for in this context~\cite{Beringer:1900zz}. Since stop LSPs in our scenario decay like leptoquarks, one can set bounds on them based on previous leptoquark searches. However, many analyses have not yet been updated to include 8~TeV data~\cite{ATLAS:2013oea,Aad:2011ch,ATLAS:2012aq, Chatrchyan:2012vza, Chatrchyan:2012st, Chatrchyan:2012sv}\footnote{For interpretation of these results for stop decays in explicit trilinear $R$-parity violation see~\cite{Evans:2012bf}.}. Searches in the top--neutrino channel, which has the same signal as a stop decaying into a top and a massless neutralino in the $R$-parity conserving MSSM with a neutralino LSP, has been updated to include the full 8~TeV dataset with preliminary results~\cite{ATLAS:2013pla,ATLAS:2013cma,CMS:2013cfa}, as has the jet--muon leptoquark search at CMS~\cite{CMS:zva}. 

The current ATLAS and CMS leptoquark analyses search for final states with opposite signed, same flavor leptons. This yields upper limits on the ${\tilde{t}}_{1}$-${\bar{\tilde t}_{1}}$ production cross section for each of the three possible flavors. The cross section upper limits from the ATLAS and CMS searches are used directly; no additional detector simulation is performed. The upper limit on the cross section is easily translated into a lower bound on the stop LSP mass, since the cross section depends only on the mass and center of mass energy and falls off steeply as the mass increases.

Although the ATLAS and CMS analyses assume branching ratios of unity to a given family, we can generalize their results to arbitrary branching ratios. This is accomplished by rescaling the expected cross section limit\footnote{For a small number of searches, the expected upper limit is not publicly available.  As these searches do not observe an excess, the observed limit is used as an approximation of the expected limit.} from each search by dividing it by the appropriate branching ratio squared. It is then compared to the calculated production cross section as a function of stop LSP mass, which yields the lower bound on the stop LSP mass from that search. For a given choice of branching ratios, the search with the strongest expected mass bound is selected. Then the observed cross section limit from that search is rescaled in the same way and, finally, compared to the calculated production cross section as a function of stop LSP mass. This yields the lower bound on the stop LSP mass. No combination of the ATLAS or CMS results is attempted.  For the case of two channels with comparable limits, such a combination might be expected to extend the stop mass limit by around 50~GeV. No special treatment of signal contamination in control regions is taken into account here, but such effects should be small for these searches. Given experimental and background uncertainties, the approximate uncertainty on a given stop lower mass bound is $\pm 50$ GeV.

\begin{figure}[h]
\includegraphics[scale=0.5]{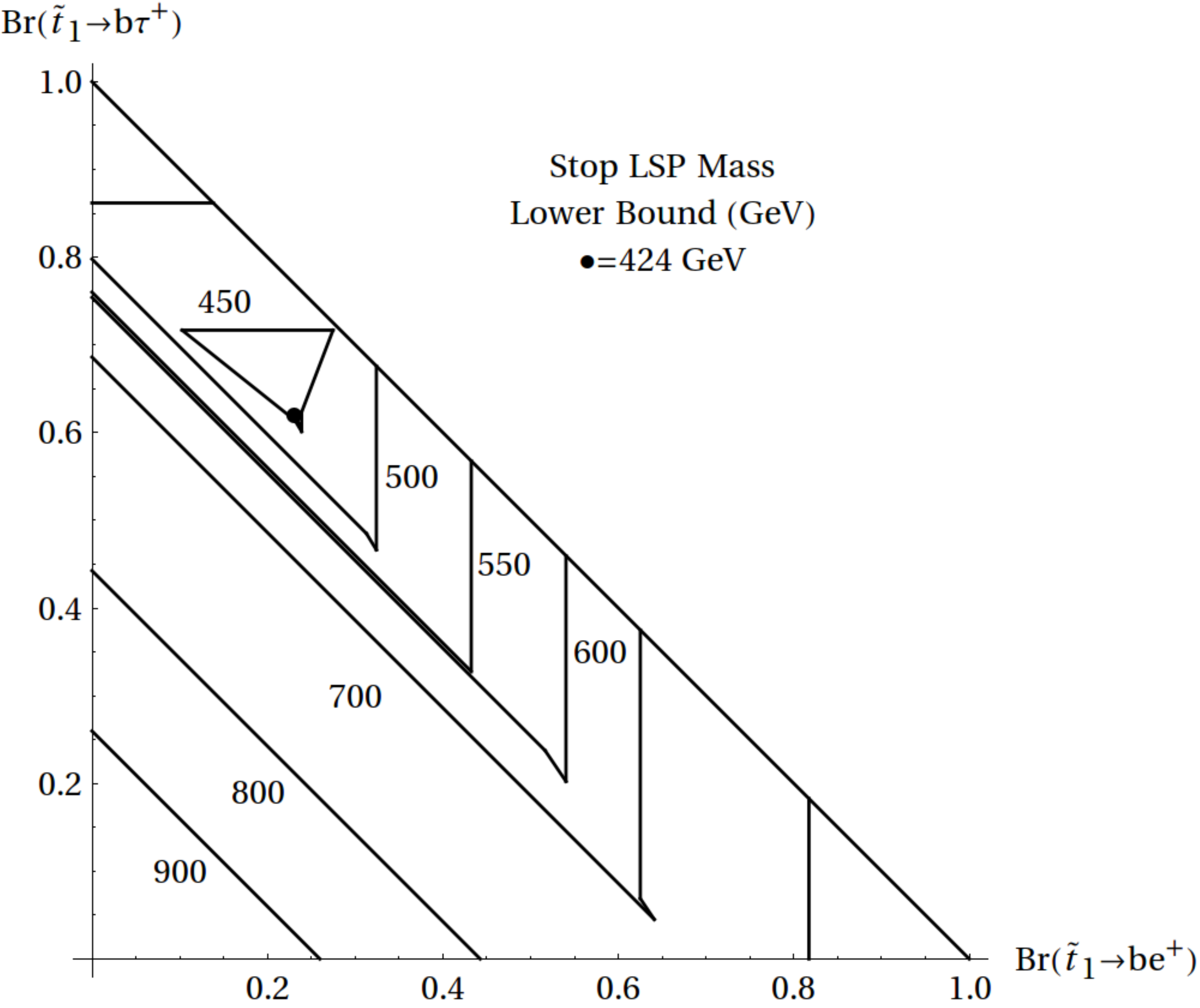}
\caption{Lines of constant stop lower bound in GeV in the $\text{Br}(\tilde t_1 \to b \, \tau^+)$ - $\text{Br}(\tilde t_1 \to b \, e^+)$ plane for an admixture stop LSP. The strongest bounds arise when the bottom--muon branching ratio is largest, while the weakest arise when the bottom--tau branching ratio is largest.}
\label{fig:stop.lower.bound}
\end{figure}

For the admixture stop LSP, the three relevant channels are the bottom--charged lepton channels. The exclusion results can, again, be plotted on a two-dimensional plot since the sum of all three branching ratios is unity. This is done in the form of lines of constant stop mass lower bound in Fig.~\ref{fig:stop.lower.bound} in the $\text{Br}(\tilde t_1 \to b  \tau^+)$ - $\text{Br}(\tilde t_1 \to b e^+)$ plane, the same plane as in Fig.~\ref{fig:Brs.central}. The absolute lowest bound, 424~GeV, occurs at $\text{Br}(\tilde t_1\to be^+)=0.23$, $\text{Br}(\tilde t_1\to b \mu^+)=0.15$, $\text{Br}(\tilde t_1\to b \tau^+)=0.62$. It is marked by a dot. The bounds are stronger in the three corners of the plot where one of the branching ratios is unity. The strongest of these three bounds corresponds to decays purely to bottom--muon. This reflects the fact that this is the easiest of the three channels to detect and the search has been performed with the most data (20 fb$^{-1}$) and at the highest energy (8 TeV). The weakest of these bounds corresponds to decays purely to bottom--tau because this channel is the hardest to detect. The contours are each composed of several connected straight line segments. The straightness of the segments is due to the fact that the bound is always coming from a single channel (the one with the strongest expected bound) and so only depends on one of the three significant branching ratios. Cross referencing Fig.~\ref{fig:stop.lower.bound} with Fig.~\ref{fig:Brs.3sigma} shows that the lowest stop mass bounds overlap the part of the normal hierarchy with a large branching ratio to bottom-tau and an inverted hierarchy with a large $\theta_{23}$ and a large branching ratio to bottom-tau.

\begin{figure}[h]
\includegraphics[scale=0.65]{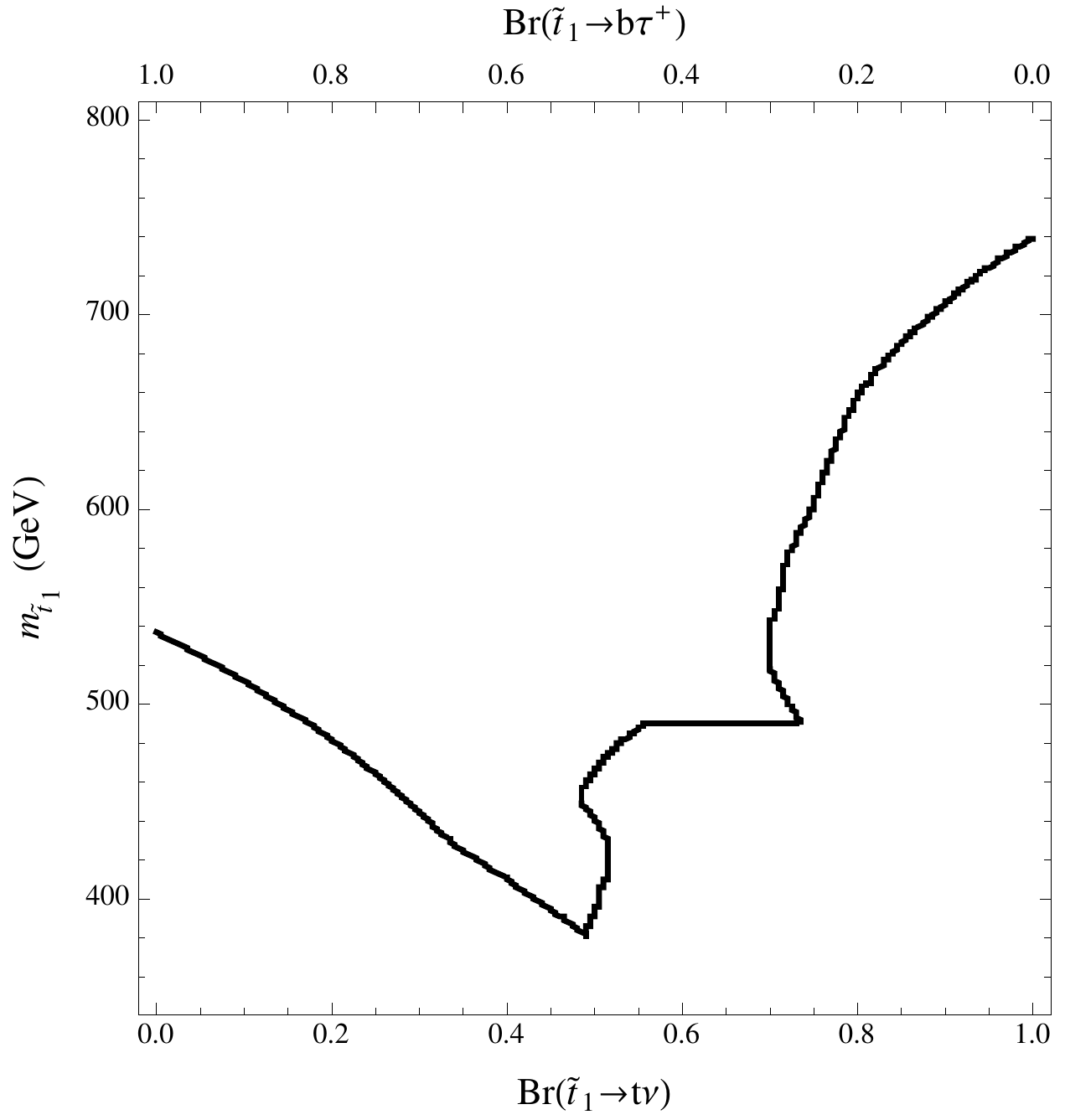}
\caption{The lower mass bound on a mostly right-handed stop--which decays predominantly into a bottom-charged lepton and a top-neutrino. It is plotted as a function of the branching ratio into top neutrino (bottom axis) and bottom--tau (top axis). The lowest allowed mass is at about 380 GeV for $\text{Br}\left(\tilde t_1 \to t \nu\right) \approx 0.5$.}
\label{fig:r.stop.lower.bound}
\end{figure}

For the right-handed stop, the production cross section limit is determined only by the stop mass and one of its branching ratios. In Fig.~\ref{fig:r.stop.lower.bound} the stop mass lower bound is plotted versus the branching ratio, with bottom--tau branching ratio on the top axis and top neutrino branching ratio on the bottom axis. Values below the plotted line are ruled out--with the exception of two pockets of allowed masses where the blue line is double valued;
for example, between $0.70 \lesssim \text{Br}(\tilde t_1 \to t \nu)\lesssim0.75$.
The lowest allowed mass is at about 380 GeV for $\text{Br}\left(\tilde t_1 \to t \nu\right) \approx 0.5$.
There is also a small allowed window, around 30 GeV wide, for the stop to have a mass similar to the top, when the branching ratio to top--neutrino dominates. This is not displayed in Fig.~\ref{fig:r.stop.lower.bound}. 

The lower bounds discussed here have the potential to be significantly improved by further analysis by the experimental groups. Some of these were mentioned above, but are listed explicitly below.  With these improvements alone and only minor re-optimization, a several hundred GeV improvement in stop mass lower bound might be obtained.
\begin{itemize}
	\item Current analyses are conducted under the assumption that a leptoquark decays dominantly into a jet and a single generation of lepton. The branching ratios in this model tend to have significant values in two or more generations (see for example Fig.~\ref{fig:Brs.3sigma}). Therefore, an analysis that takes this into account can improve the bounds by combining the bounds from different channels. This also opens the possibility of a different-flavor (e.g. electron--muon) final state, which should have strong constraints as well.\footnote{The cross-channel case when top--neutrino and bottom--tau decays dominate should not add as much, since the composition of the final state is identical to a semi-leptonic top decay.  The kinematic features may still be significantly different, e.g. with a high-transverse-momentum tau and, therefore, this channel might still be explored.}
	\item For stop LSPs, the jet accompanying the charged lepton must be a bottom quark. Therefore, an analysis with b-tagging can also help improve the bounds in the bottom--electron and bottom--muon channels by significantly reducing the jets plus $W$ or $Z$ boson background.
	\item The bottom--charged lepton channels offer an opportunity to discover the stop near the top mass.  Currently, the top--neutrino exclusion limits have an $\sim30$~GeV wide hole around the top mass.  The leptoquark limits have not been extended down to that range at the LHC, but they could be to demonstrate that the stop LSP does exist in that hole.
	\item Of course, the most straightforward improvement would come from analyzing the most up-to-date run data; that is, the 8 TeV run at 20 fb$^{-1}$.
\end{itemize}

\subsection{Sbottom LSP}
\label{sec:sbottom.lower.bounds}

In this Section, an analysis similar to that of the stop is conducted for a sbottom LSP; namely investigating its branching ratios and mass lower bound. Because many of the key points parallel the stop analysis, the discussion of both the sbottom decays and lower bound are combined here into a single short subsection.

The allowed decay channels for a sbottom LSP were given in Eq.~(\ref{PD2}). The associated partial widths are found to be
\begin{eqnarray}
\label{sbottom.width1}
\Gamma(\tilde b_1 \to b \, \nu_i)&=&\frac{1}{16\pi}(|G^L_{\tilde b_1 b\chi^0_{6+i}}|^2+|G^R_{\tilde b_1 b\chi^0_{6+i}}|^2) m_{\tilde b_1}
	\\
\label{sbottom.width2}
	\Gamma(\tilde b_1 \to t \, \ell_i^-)&=&\frac{1}{16\pi}(|G^L_{\tilde b_1 t\chi^\pm_{2+i}}|^2+|G^R_{\tilde b_1 t \chi^\pm
_{2+i}}|^2)m_{\tilde b_1}	\left(1 - \frac{m_t^2}{m_{\tilde b_1}^2} \right)
	\sqrt{1-2\frac{m_t^2}{m_{\tilde b_1}^2}+\frac{m_t^4}{m_{\tilde b_1}^4}},
\end{eqnarray}
where the $G$ parameters are given in Appendix~\ref{FR}, $\chi_{6+i}^0 = \nu_i$ and $\chi^\pm_{2+i} = \ell_i$. Both the left- and right-handed sbottom couple directly to $\tilde H_d$, which leads to the largest RPV widths. However, one can still separate the phenomenology based on the composition of the LSP sbottom. Unlike the stop LSP, a sbottom LSP can have any left--right composition while remaining the LSP. That is, the sbottom mixing angle can span the entire range $\theta_b=0^\circ-90^\circ$. Also, unlike the stop, the sbottom is expected to be mostly left-- or right--handed (that is, $\theta_b\approx 0^\circ$ or $\theta_b\approx 90^\circ$) because the off-diagonal element of the sbottom mass mass matrix is suppressed by the mass of the bottom quark (this can be seen from Eq.~\ref{theta.b}). An exception to this is when the soft masses for the third generation squark doublet, $m_{Q_3}$, and the right-handed sbottom, $m_{b^c}$, are very close (order 100 GeV for TeV scale masses and a small soft trilinear term, $a_b$, see Eq.~(\ref{theta.b})). Regardless, in the interest of being completely general, all values of the sbottom mixing angle will be considered. 

The leading order amplitudes squared for the admixture sbottom LSP, as well as the purely right-handed sbottom LSP, are approximately
\begin{align}
\label{bbnu}
	| \mathcal{A}(\tilde b_1 \to b \nu_i) |^2 & \sim
		Y_b^2 \left| {V_\text{PMSN}}_{ji} \frac{\epsilon_j}{\mu} \right|^2
	\\
\label{b1tl}
	| \mathcal{A}(\tilde b_1 \to t \ell_i^-) |^2 & \sim s_b^2 Y_b^2 \left| \frac{\epsilon_i}{\mu} \right|^2,
\end{align}
where $s_b$ is $\sin \theta_b$ and there is an implicit sum over $j$. Note that $\theta_b = 0^\circ$ ($\theta_b = 90^\circ$) corresponds to a left-handed (right-handed) lightest sbottom. The term in Eq.~(\ref{bbnu}) is independent of mixing angle since there is a contribution from both the left- and right-handed sbottoms of relatively the same size. At this order, the mostly left-handed sbottom LSP ($\theta_b\approx0^\circ$) amplitude to top--charged lepton is suppressed and one must go to the next order term
\begin{align}
	|\mathcal{A}(\tilde b_1 \to t \ell_i^-)|^2 \bigg |_{\theta_b \sim 0^\circ} & \sim Y_t^2 \left|\frac{m_{\ell_i} \, {v_L}_i}{\mu \, v_d} \right|^2.
\end{align}
From this one can conclude:
\begin{itemize}
	\item Admixture and purely right-handed sbottom  LSP: here the branching ratios to bottom--neutrino and top--charged lepton should be of the same order of magnitude. Generically, the bottom--neutrino should be somewhat larger. However, in the purely right-handed sbottom case the two branching ratios will be fairly similar.
	\item Mostly left-handed sbottom LSP: in this case, the top--charged lepton channel is suppressed by both ${v_L}_i$ and the charged lepton masses. However the decay to bottom--neutrino is not suppressed and, hence, will dominate this case.
\end{itemize}

\begin{figure}[h]
\includegraphics[scale=0.4]{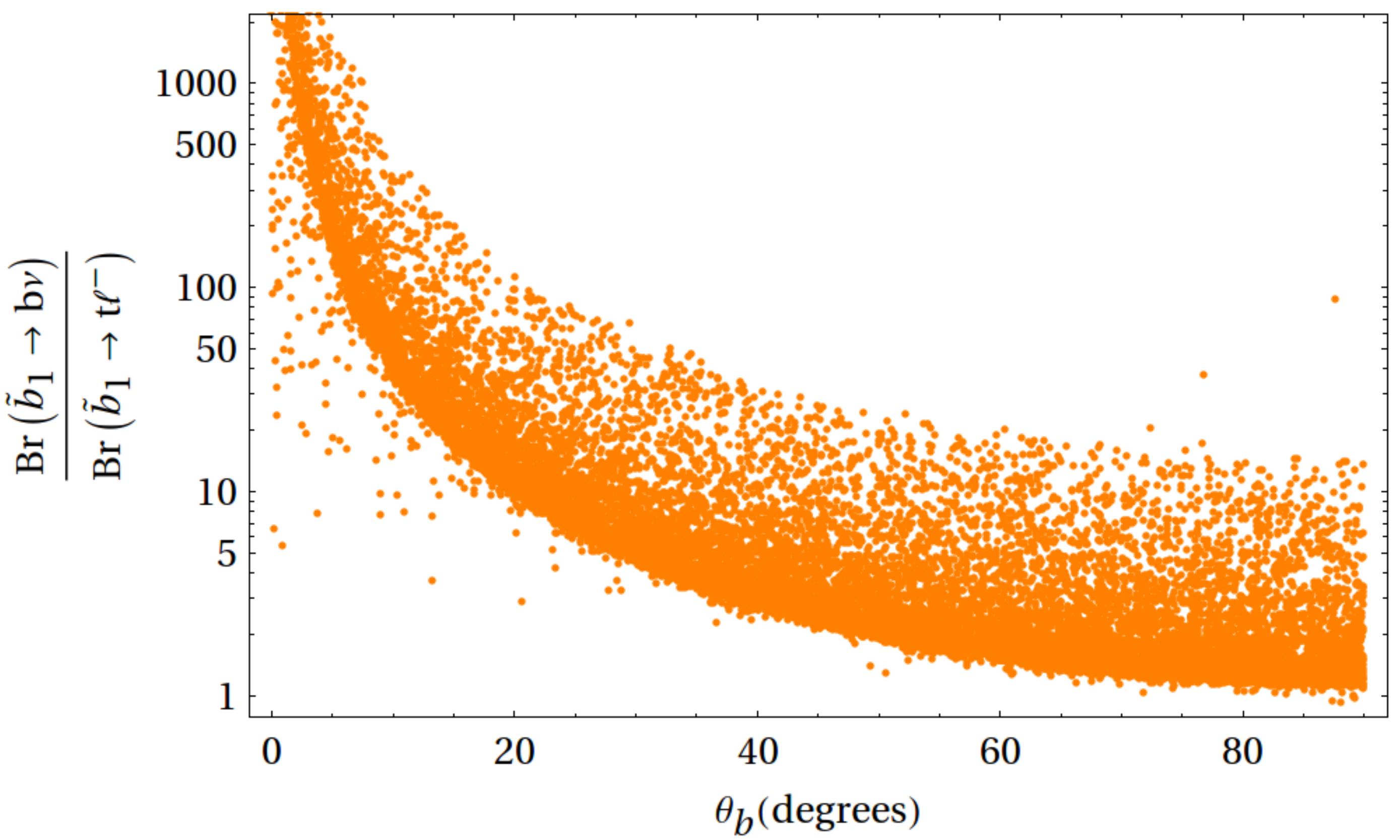}
\caption{The ratio of the branching ratio of sbottom to bottom--neutrino to the branching ratio of sbottom to top--charged lepton versus the left-right mixing angle in the sbottom sector. A $0^\circ$ ($90^\circ$) angle corresponds to a left-handed (right-handed) sbottom. Typically, one expects to be at one of the extremes of this plot as sbottom mixing is suppressed by the bottom mass.}
\label{thetab}
\end{figure}

The approximate analytic results are verified by the numerical results. These are calculated implementing the same scanning ranges as in Table~\ref{scan}, but with $\theta_t$ replaced by $\theta_b$ and $m_{\tilde t_1}$ replaced by $m_{\tilde b_1}$. The ratio $\text{Br}(\tilde b_1 \to b \nu)/\text{Br}(\tilde b_1 \to t \ell^-)$, where $\text{Br}(\tilde b_1 \to t \ell^-)\equiv\sum \limits_{i=1}^3 \text{Br}(\tilde b_1 \to t \ell_i^-)$, versus the sbottom mixing angle is displayed in Fig.~\ref{thetab}. The results closely match the approximate analytic conclusions. Sbottom lifetimes are relatively independent of the sbottom mixing angle and are typically far below the displaced vertex threshold of 1 millimeter, similar to the left-hand side of Fig.~\ref{fig:dk.length}.

We now want to produce an analogue of Fig.~\ref{fig:Brs.3sigma}. That figure was possible due to the suppressed top--neutrino channel. To produce such a figure here, where the bottom--neutrino channel is significant or even dominant, we define a new variable, the lepton branching ratio (LBr), given by

\begin{equation}
\label{LBr}
	\text{LBr}(\tilde b_1 \to t \ell_i^-) \equiv \frac{\Gamma(\tilde b_1 \to t \ell_i^-)}{\sum \limits_{i=1}^3 \Gamma(\tilde b_1 \to t \ell_i^-)} \ .
\end{equation}
This can be understood as the width of the sbottom into a single lepton generation normalized by the total width to all charged lepton generations. Note that, by definition, the three lepton branching ratios sum to unity. This allows a plot similar to Fig.~\ref{fig:Brs.3sigma} to be produced, so that one can compare the results. The sbottom situation, however, is more difficult experimentally than for the stop LSP. This is because the bottom--neutrino branching ratio  can overwhelm the top--charged lepton branching ratios to the point where they are too small to be measured at the LHC. This will be the case for the  mostly left-handed sbottom, as can be seen from Fig.~\ref{thetab}. Furthermore, here one must measure three of the four branching ratios and infer the fourth, while in the case of the admixture stop one need only measure two branching ratios to infer the third.

\begin{figure}[h]
\includegraphics[scale=0.5]{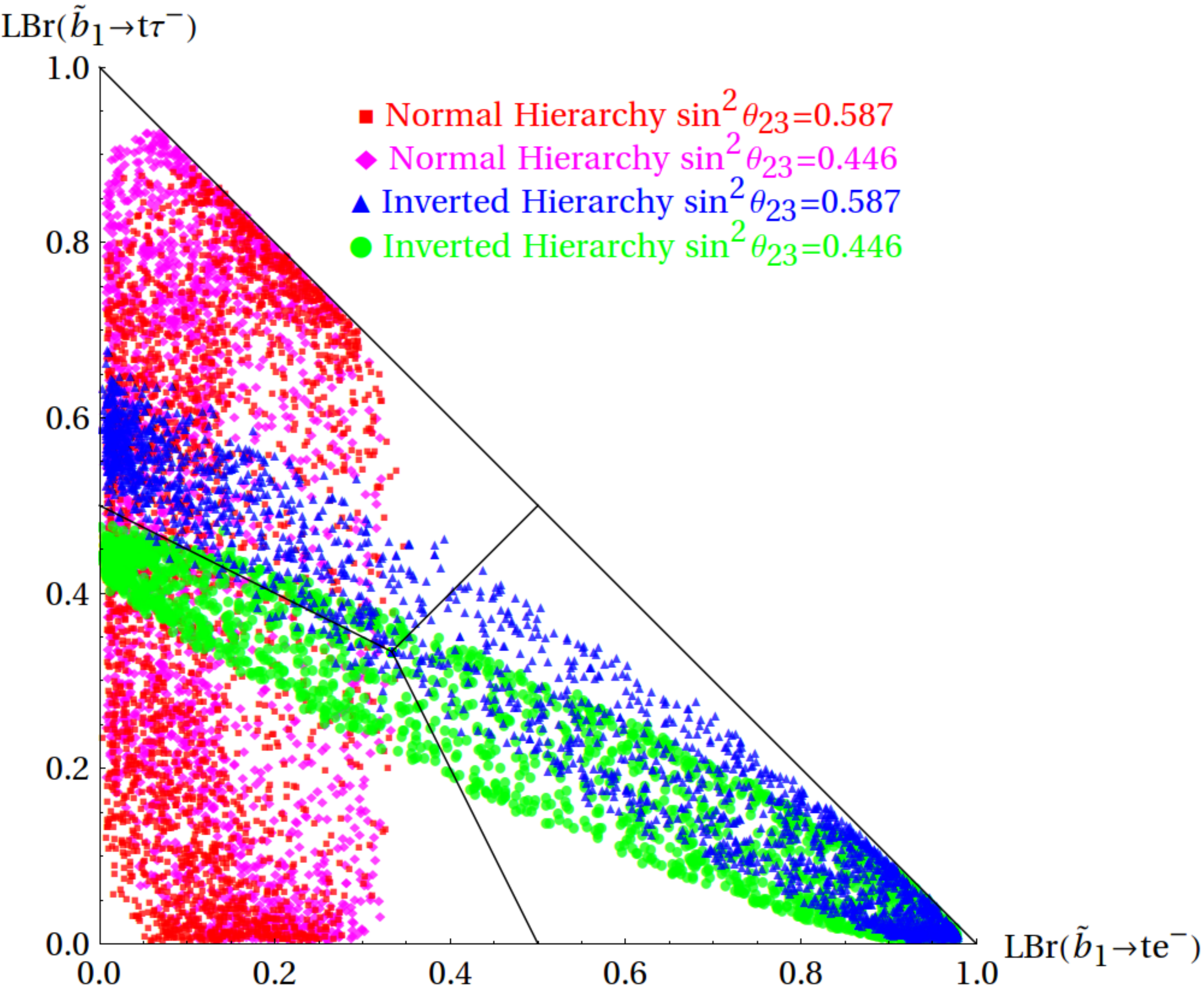}
\caption{Results of a scan over the parameters described in Table~\ref{scan}, with $\theta_t$ replaced by $\theta_b$ and $m_{\tilde t_1}$ replaced by $m_{\tilde b_1}$, are displayed in the LBr($\tilde b_1 \to t \tau)$-LBr($\tilde b_1 \to t e)$ plane where LBr is defined in Eq.~(\ref{LBr}). The details and findings of this plot are very similar to those of Fig.~\ref{fig:Brs.3sigma}.}
\label{fig:LBr}
\end{figure}

We display the lepton branching ratios in the LBrs$(\tilde b_1 \to t \tau)$-LBrs$(\tilde b_1 \to t e)$ plane in Fig.~\ref{fig:LBr}, in analogy to Fig.~\ref{fig:Brs.3sigma}. The two figures have the same features and, therefore, one can make the same conclusions as in the stop case once three of the branching ratios are measured. We will comment on this connection in the next section. In Fig.~\ref{fig:LBr} we include only points for which $\text{Br}(\tilde b_1\to b\nu)<0.99$. This excludes points where the bottom--neutrino branching ratio dwarfs the top--charged lepton branching ratio, thus making the latter unobservable. It follows from Fig.~\ref{thetab} that the plot excludes mostly left-handed sbottom LSPs. In analogy with the stop LSP case, it is preferable to base our exclusion criteria on the bottom--neutrino branching ratio instead of the mixing angle, since the former is easier to observe. Points that do not satisfy the fine-tuning criteria, Eqs.~(\ref{finetuningb}) and (\ref{finetuninga}), are excluded from Fig.~\ref{fig:LBr}.

\begin{figure}[h!]
\includegraphics[scale=0.75]{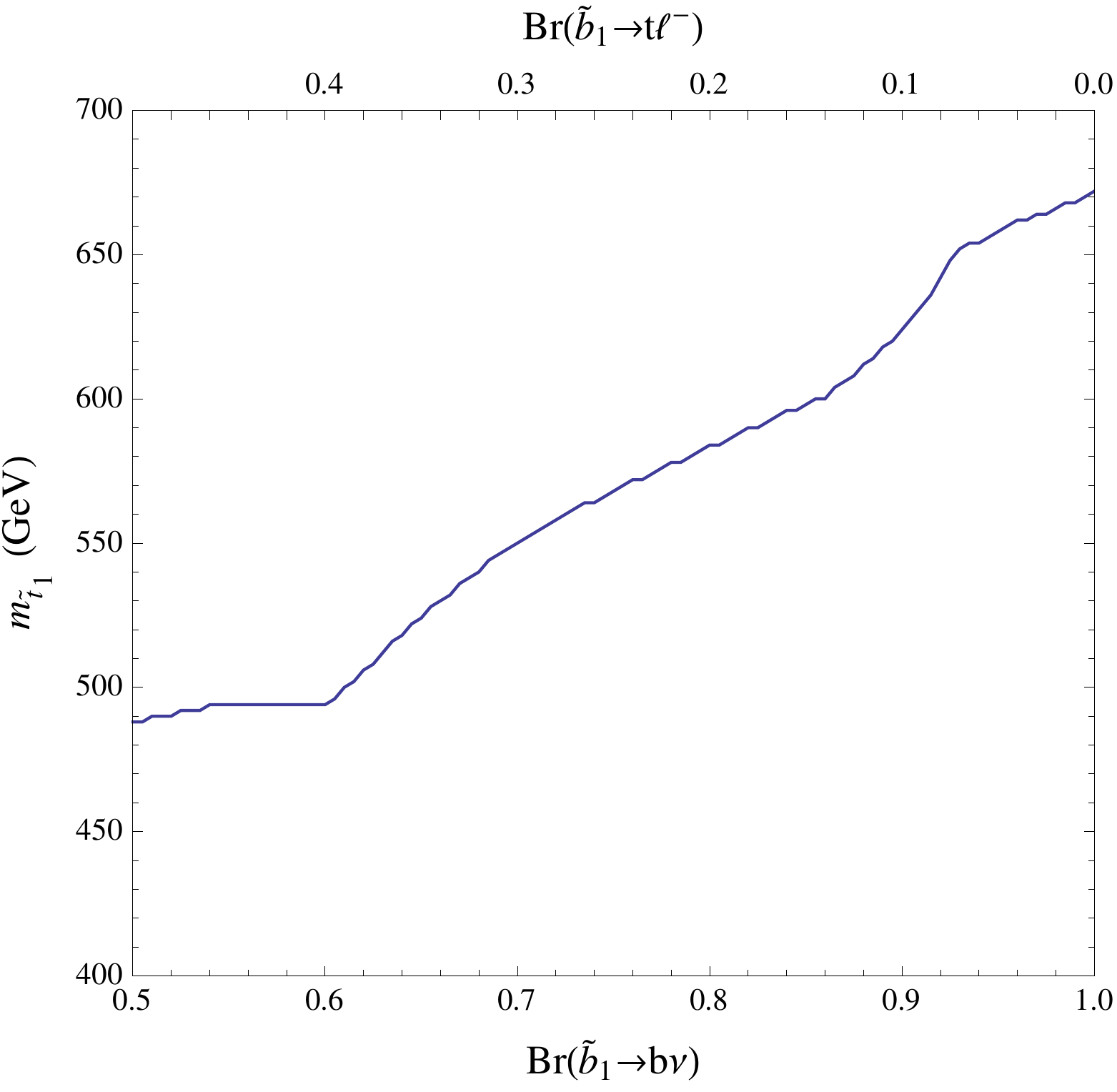}
\caption{Lower bound on the sbottom mass versus $\text{Br}\left(\tilde b_1 \to b \nu\right)$ on the bottom axis and $\text{Br}\left(\tilde b_1 \to t \ell^- \right)$ on the top axis. This bound is derived from LHC searches for the RPC decays of a sbottom to a bottom and a neutralino, reinterpreted to be our bottom-neutrino decays. }
\label{fig:b1LowerBound}
\end{figure}

In analogy to searches for the $R$-parity conserving decays of a stop into a top and a neutralino, searches have been conducted for the $R$-parity conserving decays of a sbottom into a bottom and a neutralino at both ATLAS~\cite{Aad:2013ija} and CMS~\cite{CMS:zxa} with the full 2012 data set. For massless neutralinos, these searches can be directly reinterpreted to place lower bounds on the sbottom decay to bottom-neutrino in our model, as we did for the stops in Sec.~\ref{sec:stop.lower.bounds}. These bounds are displayed in Fig.~\ref{fig:b1LowerBound} versus $\text{Br}\left(\tilde b_1 \to b \nu\right)$, which ranges in our model from 0.5 (when the sbottom is mostly right-handed) to 1 (where the sbottom is mostly left-handed), as can be seen from Fig.~\ref{thetab}. Values below the plotted line are ruled out. The stop pair production cross sections from Fig.~\ref{fig:stop.cross.section} are used for the sbottom pair production as well. This is possible since both the stop and sbottom pair production cross sections are dominantly through color interactions, and both stop and sbottom have the same color quantum number. 

Currently, there is no search which can be directly translated into lower bounds for the top-charged lepton decay channel of our sbottom, specifically searches for $t \bar t \ell^- \ell^+$ final states. However, it is possible to reinterpret current same sign lepton searches~\cite{ATLAS:2013tma,Chatrchyan:2012sa}. Such a reinterpretation will be more involved then the previous ones made in this paper and is applicable when one of the tops in our final state, $t \bar t \ell^- \ell^+$, decays leptonically as $t \to b \nu \ell^+$. The branching ratio for this top decay is about 0.1 per lepton flavor. This would produce a final state with three charged leptons,\footnote{The limits from three lepton and four lepton searches also apply, but because of the large number of disjoint signal regions, reinterpretation using these limits is better left to the LHC collaborations themselves.} two of which will have the same sign, and would therefore fall under the domain of the same sign lepton searches. However, the current bounds from the bottom-neutrino channel are relatively strong, even when that branching ratio to bottom-neutrino is only 0.5, and it is not clear whether a reinterpretation of the same sign analysis will significantly improve our present bound. We are currently investigating this issue.

\section{Discussion}
\label{disc}

One of the interesting results in this paper is the connection between the LSP decays and the neutrino hierarchy. As was shown in Figs.~\ref{fig:Brs.3sigma} and~\ref{fig:LBr}, this connection is very similar in the stop and sbottom LSP scenarios. This relationship, and the similarity, are fairly straightforward to explain and can be understood by examining the relationships in Appendix~\ref{nu} used to analyze the neutrino sector and recalling some of the analytical conclusions of the last sections. The latter of these is that the $\epsilon_i$ parameters are the dominant source of RPV and, therefore, when the decay into charged leptons is large, the amplitude to $\ell_i^\pm$ is proportional to $\epsilon_i/\mu$, see Eqs.~(\ref{t1bl}) and~(\ref{b1tl}). This yields the following approximate branching ratios and lepton branching ratios:
\begin{align}
\label{xBr}
	\text{Br}(\tilde t_1 \to b \ell^+_i) & \sim \frac{\left|\epsilon_i\right|^2}{\sum \limits_{j=1}^3 \left|\epsilon_j\right|^2}
\\
\label{xLBr}
	\text{LBr}(\tilde b_1 \to t \ell_i^-) & \sim \frac{\left|\epsilon_i\right|^2}{\sum \limits_{j=1}^3 \left|\epsilon_j\right|^2}.
\end{align}
The similarity between these two equations already explains why Figs.~\ref{fig:Brs.3sigma} and~\ref{fig:LBr} are similar.

\begin{figure}[h]
\includegraphics[scale=0.5]{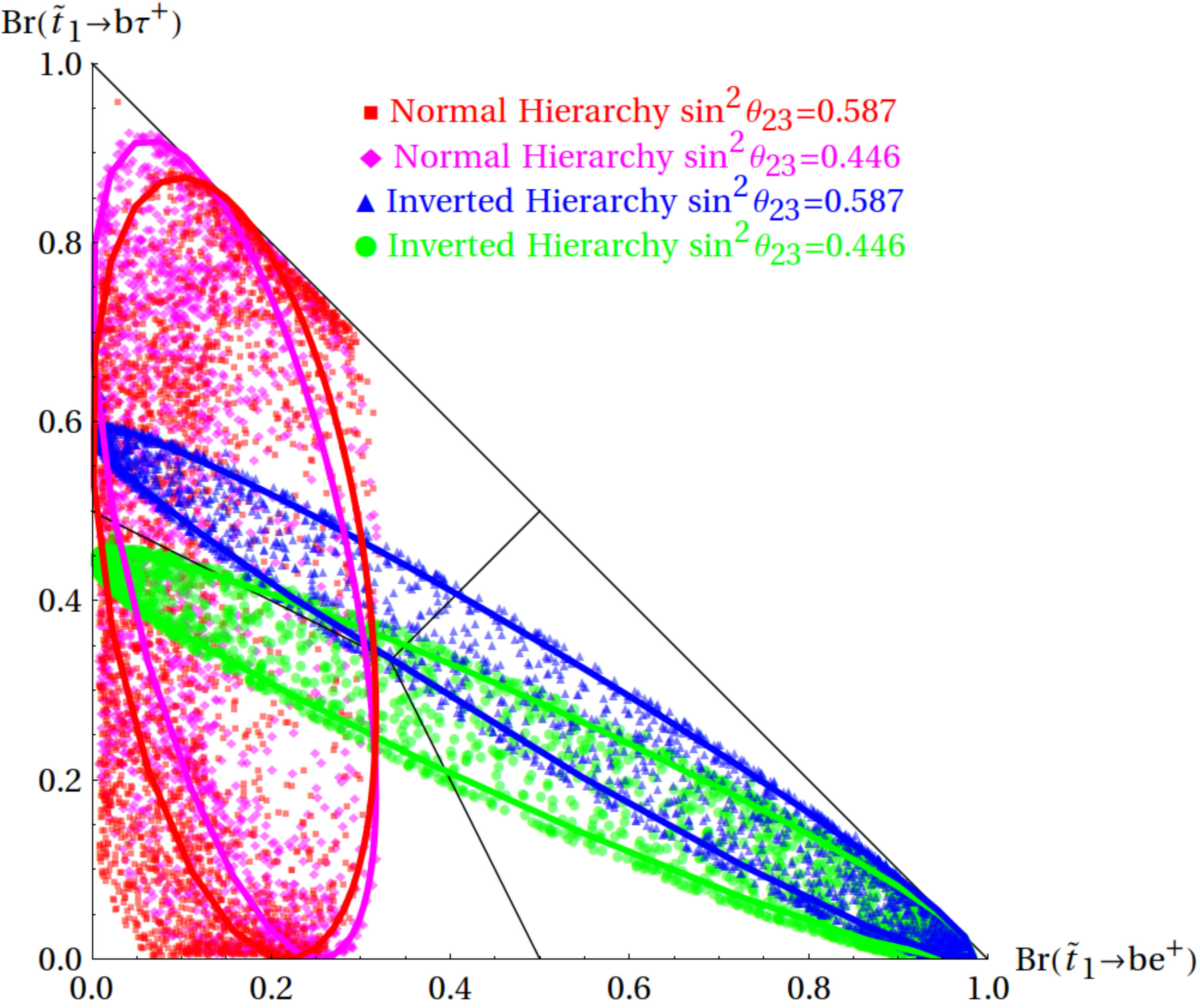}
\caption{Analytic results for the branching ratios using Eqs.~(\ref{xBr}) and (\ref{equ}) superimposed on the results from Fig.~\ref{fig:Brs.central}.}
\label{fig:analytic}
\end{figure}

The connection between the neutrino parameters and the relative sizes of $\epsilon_i$ can be qualitatively understood without appeal to random scans. Appendix~\ref{nu} relates the $\epsilon_i$ parameters to linear combinations of $E_l$ parameters weighted by the elements of the PMNS matrix,
\begin{equation}
 \epsilon_i = {V_\text{PMNS}^*}_{il} E_l.
\label{equ}
\end{equation}
Two of the $E_l$ parameters can be solved for based on the neutrino masses and mixings, but their actual values are not so important here. Let us first consider the case of a stop LSP. In the NH, $E_1=0$. Varying the relative size of $E_2$ and $E_3$ and calculating the branching ratios according to Eq.~(\ref{xBr}) traces out ellipses in the $\text{Br}(\tilde t_1\to b \tau^+)$ - $\text{Br}(\tilde t_1\to b e^+)$ plane. This can be done for both values of $\theta_{23}$. In the IH, $E_{3}=0$. Varying the relative size of $E_1$ and $E_2$ and calculating the branching ratios according to Eq.~(\ref{xBr}) again traces out ellipses in the $\text{Br}(\tilde t_1\to b \tau^+)$ - $\text{Br}(\tilde t_1\to b e^+)$ plane. This can be done for both values of $\theta_{23}$. The results, using central values for the neutrino parameters and no CP violation in the neutrino sector, are shown in Fig.~\ref{fig:analytic} superimposed over the numerical results in Fig.~\ref{fig:Brs.central}. In the case of a sbottom LSP, we find, now calculating the branching ratios using Eq.~(\ref{xLBr}), similar results with identical conclusions.

Varying the CP violating phases in the neutrino sector will move the ellipses in such a way that they fill out the same regions that were filled by the scan, thereby demonstrating the agreement between the analytic approximation and the numerical results. The same analysis would also apply to the ${v_L}_i$ parameters in cases where they dominate the decays (an example of which will be discussed shortly). The crucial features of this theory that lead to these predictions are that the $R$-parity violation is controlled by the flavorful parameters $\epsilon_i$ and ${v_L}_i$, which also give rise to neutrino masses and mixing, and that one of the neutrinos is massless.

This analytical understanding is quite powerful since it indicates that the results displayed in Fig.~\ref{fig:Brs.3sigma}, the bullet points associated with this figure and Fig.~\ref{fig:LBr} are fairly independent of many of the assumptions that have been made in this paper--which could, therefore, be relaxed or altered. These assumptions are briefly summarized here.
\begin{itemize}
	\item GUT gaugino relations: The $SO(10)$ GUT relationships for the gaugino masses has been assumed: $M_R:M_{BL}:M_2:M_3 \sim 1:1:2:5$. However, according to the analytical analysis conducted here, this would have very little impact on the relationship between the neutrino hierarchy and the branching ratios. Therefore, a bottom-up approach that does not assume this relationship would yield similar results.
	\item Gauge group: The analysis conducted in this paper was in the context of the minimal SUSY $SU(3)_C \times SU(2)_L \times U(1)_{3R} \times U(1)_{B-L}$ model. There are other gauge groups, which include a $B-L$ factor and a minimal particle content, that reduce to the SM gauge group once the $B-L$ factor is broken. They share some key features with the model studied here. These features are: all anomalies are canceled by the introduction of three right-handed neutrinos and the minimal particle content does not require a new $B-L$ Higgs sector since the right-handed sneturinos can play that role. Some examples are $SU(3)_C \times SU(2)_L \times U(1)_{Y} \times U(1)_{B-L}$~\cite{Barger:2008wn,Ambroso:2009jd} and $SU(3)_C \times SU(2)_L \times U(1)_{Y} \times U(1)_{X}$~\cite{FP:2009gr}, where $X$ is a linear combination of hypercharge and $B-L$. These common features lead to the prediction that the lightest neutrino is massless and consequently link the squark LSP decays to the neutrino hierarchy, in a similar fashion to Fig.~\ref{fig:Brs.3sigma}.
	\item Squark LSPs: Third generation squark LSPs were studied here. However, the same connection between the neutrino hierarchy and the LSP branching ratios would hold true for the first two generations as well. One difference is that the first two generations do not couple to the Higgs fields very strongly. Therefore, their dominant decay channels will be due to gauginos mixing with the neutrinos and charged leptons. This also means their lifetimes will be, on average, longer and there might be more points in parameter space with displaced vertices. Another difference is that left-right mixing angles in these generations are expected to be negligible, suppressed by the corresponding fermion mass. Therefore, one will only have the purely right- or left-handed LSPs.
	\item The parameter scan, Table~\ref{scan}: While we only scanned a finite parameter space, the analytical arguments given in this section indicate that extending the parameter space of the scan will result in similar behavior.
	\item Radiative corrections to neutrino masses: Our analysis of the neutrino sector has been carried out at tree level. However, significant radiative corrections could be present, especially when the $\epsilon_i$ parameters are relatively large. Using the results of~\cite{Diaz:2003as}, we have found that the dominant contributions carry the same flavor pattern as the tree level neutrino masses. This leaves the crucial elements discussed in this section unchanged and therefore the results. Furthermore, while the subdominant contributions do introduce a new flavor pattern, they are only significant at very large $\epsilon_i$ values. We have excluded these points from our analysis due to the resulting fine-tuning in the neutrino sector. Therefore, such radiative corrections do not effect our conclusions. See Appendix~\ref{nu} for a further discussion of this matter.
\end{itemize}

Of course, the interpretation of any leptoquark-like experimental signals in the context of Figs.~\ref{fig:Brs.3sigma} and~\ref{fig:LBr} would need further evidence in order to conclude that the model discussed in this paper corresponds to reality. Specifically, the discovery of $Z_R$ and the presence of supersymmetry should be confirmed. This latter point would probably be satisfied by the discovery of a SUSY particle beyond the LSP, which would subsequently decay into the LSP.

There exist other models with leptoquark-like signatures similar to the signatures discussed in this paper, but which do not have the same connection to neutrino physics. Such models are perhaps less well-motivated. Nevertheless, we now turn to a brief discussion of three such models. We focus, in particular, on how the theory discussed in this paper can be experimentally distinguished from these potential mimics. The most obvious example of such a model is explicit bilinear $R$-parity violation. In this case, one simply extends the $R$-parity conserving MSSM superpotential by adding the term $\epsilon_{i} L_{i} H_u$, without an understanding of the origin of this term or the suppression of baryon number violation and, hence,  proton decay. At tree-level, this model contains only one massive neutrino. It relies on radiative corrections to neutrino masses to make it consistent with experimental results~\cite{Hirsch:2000ef, Diaz:2003as}, which dictate that there are at least two massive neutrinos. The hierarchy between the tree-level and loop-level neutrino masses is consistent with the normal hierarchy, so an independent discovery of an inverted hierarchy would probably rule this model out. Also, the discovery of a heavy neutral gauge boson at the LHC would suggest our minimal SUSY $B-L$ model. Stop LSPs in this bilinear $R$-parity violating model were discussed in~\cite{Hirsch:2003fe}.

Another example is explicit lepton number violating trilinear RPV, where the superpotential includes the terms
\begin{equation}
	W \supset \lambda_{ijk} L_i L_j e^c_k + \lambda_{ijk}'  Q_i  L_j  d^c_k \ .
\end{equation}
Note that these terms are the same as the ones appearing in our theory after rotating away the $\epsilon_i$ terms, see Eq.~(\ref{eq:TRPV}). The two models differ, however, since in our theory the left-handed sneutrino attains a VEV, which effects the neutrino sector. Again, in the explicit trilinear RPV model there is no mechanism for understanding the suppression of baryon number. The  $\lambda'$ terms allow the stops and sbottoms to decay like leptoquarks. These trilinear terms also contribute to neutrino masses~\cite{Borzumati:1996hd}, but through loops diagrams that are more involved and have more freedom than in our model. This loosens the connection between neutrino masses and LSP decays. Furthermore, while $\lambda'$ allows all of the sbottom decays discussed in this paper and the decay of the stop into bottom--charged lepton, it does not allow the stop to top--neutrino channel. The top--neutrino channel, if observed, would therefore rule out $R$-parity violation dominated by trilinear terms. In addition, such terms would not, generically, be associated with a new massive neutral gauge boson.

Leaving SUSY behind, we briefly consider the phenomenology of a leptoquark addition to the SM. There are several possible types of leptoquarks, that is, with differing quantum numbers. Limiting the discussion to those that couple to matter, only leptoquarks which are doublets of $SU(2)_{L}$ do not lead to tree-level proton decay~\cite{Arnold:2013cva} and are, therefore, safe. These have the SM charges $(\mathbf{3},\mathbf{2},7/6)$ and $(\mathbf{3},\mathbf{2},1/6)$. We label the two component fields of each of these $SU(2)_{L}$ doublets as $\psi_Y^Q$, where $Y$ is the hypercharge and $Q$ is the electric charge. The mass splitting between these two component fields will be on the electroweak level and, therefore, if one is observable, there is a good chance the other should be as well. The decays of these leptoquarks are much less constrained, since the couplings that control them are relatively free. Therefore, assuming that leptoquarks couple only to the third generation, the leptoquarks have the following decays:
\begin{align}
\label{eq:LQ.1}
	& \Psi^{5/3}_{7/6} \to t \tau^+ \text{ and } \Psi^{2/3}_{7/6} \to b \tau^+,
\\ \label{eq:LQ.2}
	& \Psi^{5/3}_{7/6} \to t \tau^+ \text{ and } \Psi^{2/3}_{7/6} \to t  \bar \nu_\tau \ ,
\\
\label{eq:LQ.3}
	& \Psi^{2/3}_{1/6} \to b \tau^+ \text{ and } \Psi^{-1/3}_{1/6} \to b \bar \nu_\tau,
\end{align}
where the decays that have equal couplings due to $SU(2)_{L}$ symmetry are grouped together. Since both components of the leptoquark should be discoverable, and therefore both decays in each of the above equations, it should be possible to distinguish these from a stop or sbottom LSP. Furthermore, leptoquarks are not connected to neutrino masses and, therefore, it is not possible to predict the relative sizes of the decay channels\footnote{An exception exists when the leptoquark is embedded in a multiplet which does contribute to neutrino masses, \textit{e.g}~\cite{FileviezPerez:2008dw,Dorsner:2005fq}}. A leptoquark would also not necessarily be associated with a new neutral gauge boson and, of course, since they are scalar fields, there is a gauge hierarchy problem associated with having TeV scale leptoquarks.

\section{Summary}
\label{out}

The most minimal $B-L$ extension of the MSSM must always spontaneously break $R$-parity and, in addition, predicts the existence of a TeV scale neutral gauge boson, $Z_R$, two light sterile neutrinos and a Majorana contribution to neutrino masses coming from $R$-parity violation. Such a model is well-motivated by string theory.

This paper examined the phenomenology of third generation squark LSPs within the context of the minimal $SU(3)_{C} \times SU(2)_{L} \times U(1)_{3R} \times U(1)_{B-L}$ theory, which falls under this general class of models. Because of $R$-parity violation, these LSPs can now decay. Due to the connection between $R$-parity violation and neutrino masses, one can potentially make statements about the neutrino mass hierarchy based on the LSP branching ratios. The relevant results for the stop and sbottom LSPs are shown in Fig.~\ref{fig:Brs.3sigma} and Fig.~\ref{fig:LBr} respectively. If these quantities are measured at the LHC, their location on the plots potentially can extract information about the neutrino hierarchy. A quick summary of the conclusions made in Section~\ref{dks.nuspec} are
\begin{itemize}
	\item If the branching ratio to bottom--electron dominates, then the neutrino masses are most likely to be in the inverted hierarchy.
          \item If the branching ratio to bottom--tau dominates, then the neutrino masses are most likely in the normal hierarchy, or the inverted hierarchy with $\sin^2 \theta_{23} \sim 0.587$.
            \item A dominant branching ratio into bottom--muon would suggest either the normal hierarchy, or the inverted hierarchy with $\sin^2 \theta_{23} \sim 0.446$.
\end{itemize}
It was furthermore shown that these correlations are a result of the fact that the flavorful parameters, $v_{L_i}$ and $\epsilon_i$, simultaneously govern $R$-parity violation and the mixing in the neutrino sector.

Of course, the above conclusions depend strongly on experimental reality. Even if one discovers a particle decaying like a third generation squark LSP in this theory, to have some confidence in these conclusions, the heavy $Z_R$ gauge boson associated with this model should also be discovered. In addition,  the presence of $R$-parity violating supersymmetry must also be confirmed.
Since they require RPV, leptoquark-like decays of the stop and sbottom LSPs are not typically associated with SUSY models. Hence, they have not been as vigorously searched for as other SUSY signatures. Therefore, at this point, the bounds on these decay channels are not as strong as one might expect. In particular, a stop with the decays discussed in this paper could be as light as $\sim 424$~GeV and remain undetected, see Figs.~\ref{fig:stop.lower.bound} and \ref{fig:r.stop.lower.bound}. To highlight this, we outlined some improvements that can be made to further strengthen existing bounds in Section~\ref{sec:sbottom.lower.bounds}.

\section{Acknowledgments}
S.Spinner is indebted to P. Fileviez Perez for extensive discussions and a long term collaboration on related topics. S. Spinner would also like to thank the Max-Planck Institute for Nuclear Physics for hospitality during the early part of this work and T. Schwetz for useful discussion. B.A. Ovrut, A. Purves and S. Spinner are supported in part by the DOE under contract No. DE-AC02-76-ER-03071 and by the NSF under grant No. 1001296. The work of Z. Marshall is supported by the Office of High Energy Physics of the U.S. Department of Energy under contract DE-AC02-05CH11231.

\newpage
\appendix

\section{Neutralinos and Neutrinos:}
%
\label{nu}
$R$-parity violation allows all fermions with the same quantum numbers to mix and form physical states which are linear combinations of the original fields. In the basis $\left(\tilde W_R, \ \tilde W^0, \ \tilde H_d^0, \ \tilde H_u^0, \ \tilde B', \ \nu_3^c, \ \nu_i\right)$ with $i=1,...,3$, the neutralino mass matrix is given by
\begin{equation}
\label{neutralino}
	{\cal M}_{\chi^0} =
	\begin{pmatrix}
			M_R
		&
			0
		&
			-\frac{1}{2} \, g_{R} \, v_d
		&
			\frac{1}{2} \, g_R \, v_u
		&
			0
		&
			-\frac{1}{2} g_R v_R
		&
			0_{1 \times 3}
	\\
			0
		&
			M_2
		&
			\frac{1}{2} \, g_2 \, v_d
		&
			-\frac{1}{2} \, g_2 \, v_u
		&
			0
		&
			0
		&
			\frac{1}{2} \, g_2 \, {v_L}_i^*
	\\
			-\frac{1}{2} \, g_{R} \, v_d
		&
			\frac{1}{2} \, g_{2} \, v_d
		&
			0
		&
			-\mu
		&
			0
		&
			0
		&
			0_{1 \times 3}
	\\
			\frac{1}{2} \, g_R \, v_u
		&
			-\frac{1}{2} \, g_2 \, v_u
		&
			-\mu
		&
			0
		&
			0
		&
			0
		&
			\epsilon_i
	\\
			0
		&
			0
		&
			0
		&
			0
		&
			M_{BL}
		&
			\frac{1}{2} \, g_{BL} \, v_{R}
		&
			-\frac{1}{2} \, g_{BL} \, {v_L}_i^*
	\\
			-\frac{1}{2} g_R v_R
		&
			0
		&
			0
		&
			0
		&
			\frac{1}{2} \, g_{BL} \, v_{R}
		&
			0
		&
			\frac{1}{\sqrt 2} \, {Y_\nu}_{i3} \, v_u
	\\
			0_{3 \times 1}
		&
			\frac{1}{2} \, g_2 \, {v_L}_j^*
		&
			0_{3 \times 1}
		&
			\epsilon_j
		&
			-\frac{1}{2} \, g_{BL} \, {v_L}_j^*
		&
			\frac{1}{\sqrt 2} \, {Y_\nu}_{j3} \, v_u
		&
			0_{3 \times 3}
	\end{pmatrix},
\end{equation}
where
\begin{equation}
	\epsilon_i \equiv \frac{1}{\sqrt 2} Y_{\nu i3} v_R
\end{equation}
are the parameters of the induced bilinear $R$-parity violating terms. We have suppressed terms that are quadratic in the neutrino mass parameter, \textit{e.g.} ${v_L}_i Y_{\nu ij}$.

The neutralino mass matrix, Eq.~(\ref{neutralino}), has the schematic form
\begin{equation}
	\mathcal{M}_{\chi^0} =
	\begin{pmatrix}
		M_{\chi^0}
		&
		m_D
		\\
		m_D^T
		&
		0_{3 \times 3}
	\end{pmatrix},
\end{equation}
where $M_{\chi^0}$ is a six-by-six matrix of order a TeV and $m_D$ is six-by-three matrix of order an MeV. This allows the mass matrix to be diagonalized perturbatively. The diagonal neutralino mass matrix is
\begin{equation}
	\label{chi.diag}
	\mathcal{M}_{\chi^0}^D = \mathcal{N}^* \mathcal{M}_{\chi^0} \mathcal{N}^\dagger
\end{equation}
with
\begin{equation}
	\mathcal{N} =
	\begin{pmatrix}
		N & 0_{3 \times3}
		\\
		 0_{3\times3} & V_{PMNS}^\dagger
	\end{pmatrix}
	\begin{pmatrix}
		1_{6\times6} & -\xi_0
		\\
		\xi_0^\dagger & 1_{3 \times 3}
	\end{pmatrix},
\end{equation}
where the second matrix on the right-hand side rotates away the neutrino/neutralino mixing. This quantity is of interest since it is ultimately used in the Feynman Rules given in Appendix~\ref{FR} to calculate the third generation squark decay widths. The first matrix diagonalizes the neutralino states and the neutrino states. Equation~(\ref{chi.diag}) specifies the relationship between the gauge eigenstates, $\psi^0$, and the mass eigenstates $\chi^0$:
\begin{equation}
	\chi^0 = \mathcal{N} \psi^0,
\end{equation}
where the first six states in $\chi^0$ are the TeV scale neutralino states labeled from lightest to heaviest and the last three are the physical neutrino states.

Equation~(\ref{chi.diag}) can be used to solve for the six-by-three matrix $\xi_0$:
\begin{equation}
	\label{xi0}
	\xi_0 = M_{\chi^0}^{-1} m_D.
\end{equation}
The rows of $\xi_0$ are the gaugino gauge eigenstates and the columns correspond to the neutrino gauge eigenstates. These are explicitly labeled and presented below:
\begin{align}
	{\xi_0}_{\tilde W_R \nu_i} & = \frac{g_R \mu}{8 d_{\chi^0}}
	\left[
		2 M_{BL} v_u
		\left(
			 g_2^2 v_d v_u - 2M_2 \mu
		\right) \epsilon_i
		-g_{BL}^2 M_2 v_R^2 \left(v_d \epsilon_i + \mu {v_L}_i^* \right)
	\right]
	\\
	{\xi_0}_{\tilde W_2 \nu_i} & = \frac{g_2 \mu}{8 d_{\chi^0}}
	\left[
		2 g_R^2 M_{BL} v_d v_u^2 \epsilon_i
		+ M_{\tilde Y} v_R^2 \left(v_d \epsilon_i + \mu {v_L}_i^* \right)
	\right]
	\\
	{\xi_0}_{\tilde H_d^0 \nu_i } & = \frac{1}{16 d_{\chi^0}}
	\left[
		M_{\tilde \gamma} v_R^2 v_u \left(v_d \epsilon_i - \mu {v_L}_i^*\right)
		- 4 M_2 \mu
		\left(
			M_{\tilde Y} v_R^2 + g_R^2 M_{BL} v_u^2
		\right) \epsilon_i
	\right]
	\\
	{\xi_0}_{\tilde H_u^0 \nu_i } & = \frac{1}{16 d_{\chi^0}}
	\left[
		M_{\tilde \gamma} v_R^2 v_d \left(v_d \epsilon_i + \mu {v_L}_i^* \right)
		+ 4 g_R^2 \mu M_2 M_{BL} v_d v_u \epsilon_i
	\right]
	\\
\begin{split}
	{\xi_0}_{\tilde B' \nu_i } & = -\frac{1}{8 d_{\chi^0}}
	\left[
		g_{BL} g_R^2 M_2 \mu v_R^2 \left(v_d \epsilon_i + \mu {v_L}_i^* \right)
	\right.
	\\
	& \hspace{2cm} \left. + 2g_{BL} \mu v_u 
		 \left(
		 	\left(g_R^2 M_2 + g_2^2 M_R\right) v_d v_u - 2 M_R M_2 \mu
		 \right) \epsilon_i
	\right]
\end{split}
	\\
\begin{split}
	{\xi_0}_{\nu_3^c \nu_i } & = \frac{\mu}{8 v_R d_{\chi^0}}
	\left[
		\left(
			M_{\tilde \gamma} v_R^2 v_d v_u
			-2 g_{BL}^2 M_R M_2 \mu v_R^2
		\right){v_L}_i^*
	\right.
	\\
	& \hspace{2cm} \left.
	+ 2 M_{BL}
		 \left(
		 	M_2 \left( g_R^2 v_R^2 v_d - 4 M_R \mu v_u\right)
			+ 2 \left( g_R^2 M_2 + g_2^2 M_R\right)  v_d v_u^2
		 \right) \epsilon_i
	\right],
\end{split}
\end{align}
where
\begin{align}
	d_{\chi^0} & \equiv \frac{1}{4} M_2 M_{\tilde Y} \mu^2 v_R^2 - \frac{1}{8} M_{\tilde \gamma} \mu v_R^2 v_d v_u
	\\
	M_{\tilde \gamma} & \equiv g_R^2 g_{BL}^2 M_2 + g_2^2 g_{R}^2 M_{BL}+g_2^2 g_{BL}^2 M_R
	\\
	M_{\tilde Y} & \equiv g_R^2 M_{BL} + g_{BL}^2 M_R \ .
\end{align}

Using Eqs.~(\ref{chi.diag}) and (\ref{xi0}), or simply integrating out the heavy states, yields the neutrino mass matrix
\begin{equation}
	\label{nu.mass}
	{m_\nu}_{ij} = A {v_L}_i^* {v_L}_j^* + B \left({v_L}_i^* \epsilon_j + \epsilon_i {v_L}_j^* \right) + C \epsilon_i \epsilon_j \ ,
\end{equation}
with
\begin{align}
	A & = \frac
		{
			\mu \, M_{\tilde \gamma}
		}
		{
			2 \, M_{\tilde \gamma} v_u v_d - 4 M_2 M_{\tilde Y} \mu
		}
	\\
	B & = \frac
		{
			M_{\tilde \gamma} v_d \left( 2 M_{Z_R}^2+ g_{Z_R}^2 v_u^2\right) - 2 g_{Z_R}^2 g_{BL}^2 M_2 M_R \, \mu \, v_u
		}
		{
			4 M_{Z_R}^2 (M_{\tilde \gamma} v_u v_d - 2 M_{\tilde Y} M_2 \mu)
		}
	\\
	\begin{split}
	\label{C}
	C & = \frac
		{
				 2 g_{Z_R}^4 M_2 M_{BL} M_R \, \mu^2 v_u^2
				- g_{Z_R}^2 M_{BL} \mu
				\left(
					g_2^2 \, g_{Z_R}^2 M_R v_u^2
					+ g_R^2 M_2 \left(4 M_{Z_R}^2 + g_{Z_R}^2 v_u^2\right)
				\right) v_d v_u
		}
		{
			4 M_{Z_R}^4 \mu \left(2 M_{\tilde Y} M_2 \, \mu - M_{\tilde \gamma} v_d v_u\right)
		}
		\\
		& \quad \quad 
		-\frac
		{
			M_{\tilde \gamma}  v_d^2
		}
		{
			2 \mu \left(2 M_{\tilde Y} M_2 \, \mu - M_{\tilde \gamma} v_d v_u\right)
		} \ ,
	\end{split}
\end{align}
and where
\begin{equation}
	g_{Z_R}^2 \equiv g_{BL}^2+g_{R}^2 \ .
\end{equation}

The diagonal neutrino mass matrix is then given by
\begin{align}
\begin{split}
	\label{nu.mass.D}
	{m_\nu^D}_{ij} & = \left(V_{PMNS}^T \, m_\nu \, V_{PMNS}\right)_{ij}
\\
			& = A V_i V_j + B \left(V_i E_j + E_i V_j \right) + C E_i E_j \ ,
\end{split}
\end{align}
where
\begin{align}
	\label{vL}
	{v_L}_i & = V_l^* \, {V_{PMNS}}_{il} \ ,
	\\
	\label{epsilon}
	\epsilon_i & = E_l \, {V_{PMNS}^*}_{il} \ ,
\end{align}
and
\begin{equation}
	V_{PMNS} = 
	\begin{pmatrix}
		c_{12} c_{13}
		&
		s_{12} c_{13}
		&
		s_{13} e^{-i \delta}
		\\
		-s_{12} c_{23} - c_{12} s_{23} s_{13} e^{i \delta}
		&
		c_{12} c_{23} - s_{12} s_{23} s_{13} e^{i \delta}
		&
		c_{13} s_{23}
		\\
		s_{12} s_{23} - c_{12}  c_{23} s_{13} e^{i \delta}
		&
		-c_{12} s_{23} - s_{12}  c_{23} s_{13} e^{i \delta}
		&
		c_{13} c_{23}
	\end{pmatrix} \times \text{diag}(1, e^{i \alpha/2}, 1) \ ,
\end{equation}
with $c_{ab} (s_{ab}) = \cos \theta_{ab} (\sin \theta_{ab})$.

We note that there are a couple of complications which can make the picture given so far more involved. However these matters are not significant in the model discussed here. We addressed them one-by-one at this point.

In general, the PMNS matrix is a product of the matrix which diagonalizes the neutrino mass matrix and the corresponding matrix for the charged leptons, in analogy with the CKM matrix, see reference \cite{Cottin:2011fy} for example. In the MSSM, the charged leptons can be taken to be diagonal without loss of generality. Here, the Yukawa coupling contributions to the charged lepton masses can still be taken to be diagonal, but $R$-parity violation induces chargino-charged lepton mixing, Eq.~(\ref{x.matrix}), which leads to charged lepton-charged lepton mixing. We will show in the next section that this mixing is negligible, thereby justifying the approximation that the sole contribution to the PMNS matrix comes from the neutrino sector.

The analysis of the neutrino sector in this paper has been conducted at tree level. Radiative corrections to neutrino masses in explicit bilinear $R$-parity violation have been worked out in detail in~\cite{Hirsch:2000ef} and analyzed in~\cite{Diaz:2003as}. See also~\cite{Dedes:2006ni} in more general cases of $R$-parity violation. While our model is different from the bilinear $R$-parity violation scenario investigated in these papers, their results on radiative corrections should be approximately applicable here. Reference~\cite{Diaz:2003as} found that the dominant such contributions to the neutrino masses come from bottom-sbottom loops. These loops do not introduce new lepton flavor parameters. Therefore, they cannot change the flavor form of the neutrino mass matrix, Eq.~(\ref{nu.mass}), nor the resulting masslessness of the lightest neutrino. A massless neutrino, as well as the fact that LSP decays and neutrino masses are determined by the same flavorful parameters, $\epsilon_i$ and ${v_L}_i$, were the crucial components of our results, such as Figs.~\ref{fig:Brs.central} and~\ref{fig:LBr}, as shown in Section~\ref{disc}. Therefore our results remained unchanged with the inclusion of bottom-sbottom loops.

The next to leading order radiative contributions to the neutrino masses arise from various loops involving charged fermions and scalars such as tau-stau loops. Unlike the bottom-sbottom loops, these do introduce new lepton flavor parameters to the neutrino mass matrix (but not the LSP decays of course), such as the charged lepton Yukawa couplings. Investigating these contributions, however, indicates that they are only significant in the regime of large $\epsilon_i$ values, $\epsilon_i \gtrsim 10^{-1}$ GeV. This is a  region of fine-tuning in the neutrino sector, see Eqs~(\ref{finetuningb}) and~(\ref{finetuninga}) at the end of this section. Since such fine-tuned points were not included in our results, the tau-stau loops will not significantly alter our results.

Equations~(\ref{nu.mass.D}) - (\ref{epsilon}) can be used to solve for five of the six ${v_L}_i$ and $\epsilon_i$ parameters in terms of the the neutrino parameters, modulo two signs. The determinant of Eq.~(\ref{nu.mass}) is zero, so at tree-level there is one massless neutrino. In this case, the solutions to Eqs.~(\ref{nu.mass.D}) - (\ref{epsilon}) depend on whether the neutrino mass hierarchy is normal or inverted:

\begin{itemize}
\item{Normal Hierarchy}

In a theory with one massless neutrino, such as the one analyzed in this paper, the neutrino masses in the normal hierarchy are
\begin{equation}
	m_1 = 0 \ , \quad \quad m_2 = \sqrt{\Delta m_{21}^2} \ , \quad \quad m_3 = \sqrt{\Delta m_{31}^2} \ .
\end{equation}
Loop effects will contribute mass to the massless neutrino, but we continue in the limit where these contributions are negligible. For the normal hierarchy, Equation~(\ref{nu.mass.D}) then breaks down into the following six equations:
\begin{align}
\label{nu11}
	A V_1^2 + 2 B V_1 E_1 + C E_1^2 = & \, 0 \ ,
\\
\label{nu12}
	A V_1 V_2 + B \left(V_1 E_2 + V_2 E_1 \right)+ C E_1 E_2 = & \, 0 \ ,
\\
\label{nu13}
	A V_1 V_3 + B \left(V_1 E_3 + V_3 E_1 \right)+ C E_1 E_3 = & \, 0 \ ,
\\
\label{nu23}
	A V_2 V_3 + B \left(V_2 E_3 + V_3 E_2 \right) + C E_2 E_3 = & \, 0 \ ,
\\
\label{nu22}
	A V_2^2 + 2 B V_2 E_2 + C E_2^2 = &\,  m_2 \ ,
\\
\label{nu33}
	A V_3^2 + 2 B V_3 E_3 + C E_3^2 = & \ m_3 \ .
\end{align}
Equations.~(\ref{nu11}) - (\ref{nu13}) force $V_1, E_1 = 0$. The remaining system of equations, (\ref{nu23}) - (\ref{nu33}), can be solved for with respect to $E_3$:
\begin{align}
	E_2 & = \zeta_1
		\sqrt
		{
			-\frac{m_2}{m_3}
			\left
			(
				E_3^2 + \frac{A m_3}{R}
			\right)
		} \ ,
\label{eq:826}
\\
\label{eq:V2}
V_2 & = \frac{1}{A}
		\left(
			-B E_2 + \zeta_2   \sqrt{R \left(E_2^2 + \frac{A m_2}{R} \right)}
		\right) \ ,
\\
	V_3 & = \frac{1}{A}
\label{eq:V3}
		\left(
			-B E_3 + \zeta_3   \sqrt{R \left(E_3^2 + \frac{A m_3}{R}\right)}
		\right) \ ,
\end{align}
where
\begin{equation}
	R \equiv B^2 - A C
\end{equation}
and $\zeta_1$, $\zeta_2$ and $\zeta_3$ are the usual sign factors ($\pm1$) associated with solving a quadratic equation. These sign factors, however, are not all independent. They are related by
\begin{equation}
	\zeta_2= \zeta_1 \zeta_3
	\frac{
			\sqrt{-\frac{m_2}{m_3} R E_3^2} \, \sqrt{R \left(E_3 +\frac{A m_3}{R}\right)}
		}
		{
			R E_3 \sqrt{-\frac{m_2}{m_3} \left(E_3^2 + \frac{A m_3}{R} \right)}
		} \ .
\end{equation}
Inverting Eqs.~(\ref{vL}) and~(\ref{epsilon}) translates these solutions in $E_i$ and $V_i$ to $\epsilon_i$ and ${v_L}_i$.

Using Eqs.~(\ref{epsilon}) and~(\ref{eq:826}), $E_3$ can be expressed in terms of any one of the $\epsilon_i$. This is advantageous because the $\epsilon_i$ are more transparently related to stop decay branching ratios and are the more fundamental parameters in the Lagrangian. This allows one to specify one of the $\epsilon_i$ as the input parameters. Substituting Eq.~(\ref{eq:826}) and $E_1=0$ into Eq.~(\ref{epsilon}) and squaring it yields a quadratic equation for $E_3$. It is solved by
\begin{equation}
	E_3 =
	\frac
	{
		\epsilon_iV_{PMNS_{i3}}^*
		+\zeta_0
		\sqrt
		{
			- \frac{m_2}{m_3} (V_{PMNS_{i2}}^*)^2 \epsilon_i^2
			-\frac{A m_2}{R} (V_{PMNS_{i2}}^*)^2
			\left(
				(V_{PMNS_{i3}}^*)^2 + \frac{m_2}{m_3} (V_{PMNS_{i2}}^*)^2
			\right)
		}
	}
	{
		(V_{PMNS_{i3}}^*)^2+\frac{m_2}{m_3}(V_{PMNS_{i2}}^*)^2
	}.
\label{eq:439}
\end{equation}
This introduces a new sign $\zeta_0=\pm1$ into the procedure, as well as a new constraint on the sign variables. Substituting Eq.~(\ref{eq:826}) into Eq.~(\ref{epsilon}) yields
\begin{equation}
	\zeta_1 =
		\frac
		{
			(\epsilon_i-V_{PMNS_{i3}}^*E_3)
		}
		{
			\sqrt
			{
				-\frac{m_2}{m_3} \left(E_3^2 + \frac{A m_3}{R} \right)
			}
			V_{PMNS_{i2}}^*
		}.
\end{equation}
The result is that specifying the SUSY and $B-L$ parameters, as well as any one of the $\epsilon_i$ and the two signs $\zeta_0$ and $\zeta_2$, specifies the $v_{L_i}$ and the other two $\epsilon_i$.

\item Inverted Hierarchy

The neutrino masses in the inverted hierarchy are
\begin{equation}
	m_1 = \sqrt{\Delta m_{31}^2} \ , \quad \quad m_2 = \sqrt{\Delta m_{31}^2+\Delta m_{21}^2 \ }, \quad \quad m_3 = 0 \ .
\end{equation}
In this case, the procedure above is modified in the following ways:  
$m_1 \leftrightarrow m_3$, $E_1 \leftrightarrow E_3$, $V_1 \leftrightarrow V_3$. Thus, solving for $V_i$ and $E_i$ one obtains 
$V_3, E_3=0$ and the solutions above with the appropriate substitutions.

\end{itemize}

In both the normal and inverted neutrino hierarchies, since the dimensionful parameters $\epsilon_i$ are responsible for neutrino masses, there is a relationship between their overall scales. We understand this in terms of two fine-tuning criteria, and use it to inform our choice of the range of $\epsilon_i$ in our scans defined in Table~\ref{scan}. We then use these fine-tuning criteria to exclude finely tuned points from Figs.~\ref{fig:Brs.central}, \ref{fig:Brs.3sigma}, and \ref{fig:LBr}. Relaxing these criteria does not significantly change the trends displayed in those figures. In the normal hierarchy, the first criterion is that the last terms on the left hand sides of Eqs.~(\ref{nu22}),(\ref{nu33}) should not be much bigger than the right hand sides. Were they to be, this would require a delicate cancellation between the terms on the left hand sides to produce the correct neutrino masses. Specifically, the criterion is
\begin{eqnarray}
	|CE_i^2|<10\cdot m_i \ ,
	\label{finetuningb}
\end{eqnarray}
where $i=2,3$. The second criterion is that none of the $\epsilon_i$ should be much smaller than the $E_i$, since the former are just linear combinations of the latter. That is, take
\begin{eqnarray}
	10\cdot |\epsilon_i|>|E_j|
	\label{finetuninga}
\end{eqnarray}
for all $i=1,2,3$ and $j=2,3$. In the invented hierarchy, these conditions are the same except with the appropriate replacements: $m_1 \leftrightarrow m_3$, $E_1 \leftrightarrow E_3$, $V_1 \leftrightarrow V_3$.

\section{Charginos and Charged Leptons:}
\label{charginos}
The charginos mix with the charged leptons due to $R$-parity violation. The chargino mass matrix, in the basis $\left(\tilde W^+, \ \tilde H_u^+,e^c_i, \ \tilde W^-, \ \tilde H_d^-, \  e_i \right)$, is given by
\begin{equation}
\label{chargino}
{\cal M}_{{\tilde \chi}^{\pm}}	=	
	\left(
	\begin{array}{cc}
		0_{5\times5}
	&
		\mathcal{X}^T
	\\
		\mathcal{X}
	&	0_{5\times5}
	\end{array}
	\right),
\end{equation}
with
\begin{equation}
\label{x.matrix}
	\mathcal{X} =
	\begin{pmatrix}
			M_2
		&
			\frac{1}{\sqrt{2}} g_2 v_u
		&
			0
		&
			0
		&
			0
	\\
			\frac{1}{\sqrt{2}} g_2 v_d
		&
			\mu
		&
			-\frac{{v_L}_1}{v_d} m_e
		&
			-\frac{{v_L}_2}{v_d} m_\mu 
		&
			-\frac{{v_L}_3}{v_d} m_\tau
	\\
			\frac{1}{\sqrt{2}} \ g_2  {v_L}_1^*
		&
			-\epsilon_1
		&
			m_e
		&
			0
		&
			0
	\\
			\frac{1}{\sqrt{2}} \ g_2  {v_L}_2^*
		&
			-\epsilon_2
		&
			0
		&
			m_\mu
		&
			0
	\\
			\frac{1}{\sqrt{2}} \ g_2  {v_L}_3^*
		&
			-\epsilon_3
		&
			0
		&
			0
		&
			m_\tau
	\end{pmatrix}
\end{equation}
This has the schematic form
\begin{equation}
	\label{X.symb}
	\mathcal{X} = 
	\begin{pmatrix}
		X & \Gamma
		\\
		G^T & {m_\ell}_i
	\end{pmatrix},
\end{equation}
where $X$ is on the order of the SUSY soft mass scale and $\Gamma$, $G$ are proportional to RPV and, therefore, much smaller. The chargino mass matrix is diagonalized as
\begin{equation}
	\mathcal{X}^D = \mathcal{U}^* \mathcal{X} \mathcal{V}^\dagger,
\end{equation}
where $\mathcal{V}$ diagonalizes the positively charged charginos and $\mathcal{U}$ the negatively charged charginos. The relationships between the gauge eigenstates, $\psi^\pm$, and the mass eigenstates, $\chi^\pm$, are
\begin{align}
	\chi^- & = \mathcal{U} \psi^-,
\\
	\chi^+ & = \mathcal{V} \psi^+.
\end{align}
The first two components of the mass eigenstates are the physical chargino TeV scale states and the last three are the physical charged lepton states.

As with the neutralinos, the chargino/charged lepton mixing can be perturbatively rotated away. The mixing matrix that does this is used in the Feynman rules in Appendix~\ref{FR} to calculate the decay widths for the third generation squarks. Following a similar procedure as for the neutralinos, the negative chargino mixing matrix is
\begin{equation}
	\label{U}
	\mathcal{U} = 
	\begin{pmatrix}
		U & 0_{2\times3}
		\\
		0_{3\times2} & U_\ell 
	\end{pmatrix}
	\begin{pmatrix}
		1_{2\times2} & -\xi_-
		\\
		\xi_-^\dagger & 1_{3\times3}
	\end{pmatrix} \ ,
\end{equation}
where $U_\ell$ is the matrix which diagonalizes the charged lepton mass matrix. As mentioned in the previous Appendix, we will show that this matrix is approximately unity below.

Successful, perturbative, diagonalization requires
\begin{equation}
	\label{xi}
	\xi_- = - \left(X^T\right)^{-1} G.
\end{equation}
Technically, the rows of $\xi_-$ are the negative chargino gauge eigenstates and the columns are the charged lepton gauge eigenstates. However, the latter are very close to the mass eigenstates and will, therefore, be labeled accordingly:
\begin{align}
	\left(\xi_-\right)_{ \tilde W^- \ell_i} & = -\frac{g_2}{\sqrt 2 d_{X}} \left(v_d \epsilon_i + \mu {v_L}_i^*\right)
	\\
	\left(\xi_-\right)_{\tilde H_d^- \ell_i } & = \frac{1}{2 d_{X}} \left(2M_2 \epsilon_i + g_2 v_u {v_L}_i^*\right),
\end{align}
where
\begin{equation}
	d_{X} = M_2 \mu - \frac{1}{2} g_2^2 v_d v_u
\end{equation}
is the determinant of $X$.

The positive chargino mixing matrix is
\begin{equation}
	\mathcal{V} = 
	\begin{pmatrix}
		V & 0_{2\times3}
		\\
		0_{3\times2} & 1_{3\times3}
	\end{pmatrix}
	\begin{pmatrix}
		1_{2\times2} & -\xi_+
		\\
		\xi_+^\dagger & 1_{3\times3}
	\end{pmatrix} \ .
\end{equation}
Solving from diagonalization yields
\begin{equation}
	\xi_+ = -\left(X \right)^{-1} \Gamma ,
\end{equation}
where the components of $\xi_+$ are
\begin{align}
	\label{eq:wino.mixing}
	\left(\xi_+ \right)_{ \tilde W^+ \ell_i} & = -\frac{1}{\sqrt 2 d_X} g_2 \tan \beta  \, {m_\ell}_i {v_L}_i
\\
	\left(\xi_+ \right)_{ \tilde H_u^+ \ell_i} & = \frac{1}{d_X}\frac{M_2 {m_\ell}_i {v_L}_i}{v_d} \ .
\end{align}

In Appendix~\ref{nu}, it was stated that the charged lepton mixing is negligible in this model. This will be shown here at the first order of the perturbative expansion. To begin, let us examine the mass matrix squared for the negative charginos, $\chi^-$:
\begin{equation}
	M_{\chi^-} = \mathcal{X} \mathcal{X}^\dagger = 
	\begin{pmatrix}
		X X^\dagger + \Gamma \Gamma^\dagger & X G^* + \Gamma m_{\ell_i}
		\\
		G^T X^\dagger + m_{\ell_i} \Gamma^\dagger & m_{\ell_i}^2 + G^T G^*
	\end{pmatrix},
\end{equation}
where we have used the symbolic Eq.~(\ref{X.symb}). Furthermore, the $m_{\ell_i}$ terms in the one-two and two-one elements are negligible compared to the SUSY scale $X$. The matrix $M_{\chi^-}$ is diagonalized as
\begin{equation}
	M_{\chi^-}^D = \mathcal{U}^* M_{\chi^-} \mathcal{U}^T.
\end{equation}
Using Eq.~(\ref{U}), the two-two element of $M_{\chi^-}^D$ is
\begin{equation}
	\label{xi.D.22}
	(M_{\chi^-}^D)_{22} = U_\ell^* \left(\xi_-^T X X^\dagger \xi_-^* + G^T X^\dagger \xi_-^* + \xi_-^T X G^* + m_{\ell_i}^2 + G^T G^*\right) U_\ell^T.
\end{equation}
Interestingly, using the solution for $\xi_-$, Eq.~(\ref{xi}), leads to a cancellation in Eq.~(\ref{xi.D.22}) so that it simplifies to
\begin{equation}
	\label{xi.D.22.simp}
	U_\ell^* \left(m_{\ell_i}^2\right) U_\ell^T.
\end{equation}
Since $m_{\ell_i}^2$ is already diagonal, $U_\ell$ is simply the identity at the level of this approximation. One can do a similar analysis with the $\chi^+$ mass matrix and we have checked that this approximation is numerically valid thereby justifying the sole contribution to the PMNS matrix from the neutrino sector.

\section{Squarks}
\label{sfermions}

In a general SUSY scenario, all six up-type squarks mix with each other and all six down-type squarks mix with each other as well. However, flavor physics dictates that there should be little mixing between the first and second generations. Furthermore, left-right mixing in a given generation is suppressed by the corresponding fermion mass. Therefore, it is generally assumed that significant mixing only exists in the third generation, as assumption adopted in this paper as well. The sfermion masses have different $D$-term contributions in this model than in the MSSM and are therefore presented here. The mass matrices $\mathcal{M}_{\tilde t}^2$ and $\mathcal{M}_{\tilde b}^2$, in the basis $\left(\tilde t, \  {\tilde t}^{c*} \right)$ and $\left(\tilde b, \  {\tilde b}^{c*} \right)$, are
\begin{eqnarray}
\mathcal{M}_{\tilde t}^2&=&\left(\begin{array}{cc}
		m_{\tilde Q_3}^2
		+ m_{t}^2
		+ \frac{1}{2} c_W^2 c_{2\beta} M_Z^2 
		+ \frac{1}{6} s_{R}^2 M_{Z_R}^2
		&
		m_t \left(A_t - \frac{\mu}{\tan \beta} \right)
	\\
		m_t \left(A_t - \frac{\mu}{\tan \beta} \right)
		&
		m_{\tilde t^c}^2
		+ \ m_{t}^2
		+\left(\frac{1}{2}  - \frac{2}{3} s_{R}^2\right) M_{Z_R}^2
	\end{array}\right),
\label{eq_stopmassmatrix}
\end{eqnarray}	
\begin{eqnarray}	
\mathcal{M}_{\tilde b}^2 &=& \left(\begin{array}{cc}
		m_{\tilde Q_3}^2
		+ m_{b}^2
		- \frac{1}{2} c_W^2 c_{2\beta} M_Z^2 
		+ \frac{1}{6} s_{R}^2 M_{Z_R}^2
		&
		m_b \left(A_b - \tan \beta \, \mu \right)
	\\
		m_b \left(A_b - \tan \beta \, \mu \right)
		&
		m_{\tilde b^c}^2
		\ + \ m_{b}^2
		+\left(-\frac{1}{2}  + \frac{1}{3} s_{R}^2\right) M_{Z_R}^2
	\end{array}\right),
\label{eq_sbottommassmatrix}
\end{eqnarray}
where $c_{2\beta} \equiv \cos 2\beta$, $c_W \equiv \cos \theta_W$, $\theta_W$ is the weak mixing angle and $s_R \equiv \sin \theta_R = g_{BL}/\sqrt{g_{BL}^2 +g_R^2}$. This latter quantity is technically a free parameter from a low energy perspective.  However,  in the UV physics discussed in reference~\cite{Ovrut:2012wg}, it takes the value $s_R^2 \sim 0.6$. In this paper, the numerical work was carried out by scanning over the physical masses of the squarks and, therefore, this parameter is not used. Here, $m_t, \ m_b$ are the top and bottom masses and $Y_t A_t$, $Y_b A_b$ are the trilinear $a$-terms.The physical states are related to the gauge states by
\begin{align}
\label{eq_squarkmixing}
	\begin{pmatrix}
		\tilde f_1
		\\
		\tilde f_2
	\end{pmatrix}
	& =
	\begin{pmatrix}
		\cos \theta_{ f}
		&
		\sin \theta_{ f}
		\\
		- \sin \theta_{ f}
		&
		\cos \theta_{ f}
	\end{pmatrix}
	\begin{pmatrix}
		\tilde f
		\\
		\tilde f^{c*}
	\end{pmatrix} \ ,
\end{align}
where $\tilde f$ represent either $\tilde t$ or $\tilde b$ and $m_{\tilde f_1} < m_{\tilde f_2}$. The lightest sfermion is purely left-handed (right-handed) when its mixing angle is 0$^\circ$ (90$^\circ$). The mixing angles are given by
\begin{align}
	\tan 2 \theta_t & = \frac{2 m_t \left(A_t - \frac{\mu}{\tan \beta}\right)}
	{
		m_{\tilde Q_3}^2 + \frac{1}{2} c_W^2 c_{2\beta} M_Z^2
		- m_{\tilde t^c}^2 + \left(-\frac{1}{2}  + \frac{5}{6} s_{R}^2\right) M_{Z_R}^2
	},
\\
\label{theta.b}
	\tan 2 \theta_b & = \frac{2 m_b \left(A_b - \mu \tan \beta\right)}
	{
		m_{\tilde Q_3}^2 - \frac{1}{2} c_W^2 c_{2\beta} M_Z^2
		- m_{\tilde b^c}^2 + \left(\frac{1}{2}  - \frac{1}{6} s_{R}^2\right) M_{Z_R}^2
	},
\end{align}
when ${\mathcal{M}_{\tilde t}^2}_{11} > {\mathcal{M}_{\tilde t}^2}_{22}$ and ${\mathcal{M}_{\tilde b}^2}_{11} > {\mathcal{M}_{\tilde b}^2}_{22}$. When ${\mathcal{M}_{\tilde t}^2}_{11} < {\mathcal{M}_{\tilde t}^2}_{22}$, $\theta_t$ is shifted by $-\pi/2$ and when ${\mathcal{M}_{\tilde b}^2}_{11} < {\mathcal{M}_{\tilde b}^2}_{22}$, $\theta_b$ is shifted by $-\pi/2$.

It is worthwhile to note that a purely left-handed lightest stop ($\theta_t=0$) cannot be the LSP. This is because both the left-handed stop and the left-handed sbottom get some of their mass from the $m_{\tilde Q_3}^2$ soft mass parameter (as shown in Eqs.~(\ref{eq_stopmassmatrix}),(\ref{eq_sbottommassmatrix})) and their respective fermion masses, $m_t$ and $m_b$. Since $m_t^2>m_b^2$, $m_{\tilde t_1}^2>m_{\tilde b_1}^2$ for a purely left-handed lightest stop. It is possible that mixing in the sbottom sector could change this, but those effects are expected to be small since they are proportional to $m_b$ (see the off-diagonal elements of Eq.~(\ref{eq_sbottommassmatrix})). For a mostly left-handed stop ($\theta_t\approx 0$), the lightest stop can be the LSP for certain values of some parameters that do not effect the physics studied in this paper.

\section{Feynman Rules}
\label{FR}

In this Appendix, the Feynman rules for the interactions between third generation squarks, quarks and neutralinos, and charginos are listed in the physical basis. The physical neutralinos and charginos are labeled by the subscript $n$. For the neutralinos, $\chi^0_n = (\chi_1,..., \chi_6, \nu_i)$ where the first six states are the TeV scale neutralinos and the last three states are the physical neutrinos labeled by $i$. For the charginos $\chi_n^\pm = (\chi_1^\pm, \chi_2^\pm, \ell_i)$ where the first two states are the TeV scale charginos and the lass three states are the charged leptons labeled by $i$. In this case, the physical $i^\text{th}$ neutrino is given by $\chi^0_{6+i}$ and the physical $i^\text{th}$ charged lepton is $\chi^\pm_{2+i}$.

The Feynman rule for each process will be followed by an approximation of that Feynman rule relevant for the $R$-parity violating decays discussed in the paper; namely, leptoquark-like decays. This approximation will be given in the limit $M_{Z_R}^2\gg m_\text{soft}^2 \gg v_{d,u}^2$ using the perturbative diagonalizations presented in Appendices~\ref{nu} and~\ref{charginos}. We also employ the fact that $\epsilon_i^2 \gg {v_L}_i^2$ in general. This is useful for an analytic understanding of the strengths of the different decay channels.
\subsection{Stops}
For the lightest stop vertex $\tilde t_1 \, t \, \tilde \chi_{n}^0:$
\begin{equation}
	 \quad  g_{\tilde t_1 t \chi_n^0} =  G^L_{\tilde t_1 t \chi_n^0} P_L + G^R_{\tilde t_1 t \chi_n^0} P_R \ ,
\end{equation}
where
\begin{align}
	G^L_{\tilde t_1 t \chi_n^0} & = 	\frac{1}{\sqrt 2} g_R s_{\theta_t} \mathcal{N}_{n 1}^*
							+ \frac{1}{3 \sqrt 2} g_{BL} s_{\theta_t} \mathcal{N}_{n 5}^*
							- Y_t c_{\theta_t} \mathcal{N}_{n 4}^* \ ,
	\\
	G^R_{\tilde t_1 t \chi_n^0} & = 	-\frac{1}{\sqrt 2} g_2 c_{\theta_t} \mathcal{N}_{n 2}
								- \frac{1}{3 \sqrt 2} g_{BL} c_{\theta_t} \mathcal{N}_{n 5}
								-Y_t s_{\theta_t} \mathcal{N}_{n 4} \ .
\end{align}
and $P_{\frac{L}{R}}=\frac{1}{2}(1\pm \gamma_{5})$. For the neutrino components of the physical neutralinos, $\chi_{6+i} = \nu_i$, these $G$ parameters are approximated by
\\
\begin{align}
\begin{split}
	G^L_{\tilde t_1 t \nu_i} \approx
	\left(V_{PMNS}\right)_{ji}
	&
	\left[
		\frac{1}{\sqrt 2} g_R s_{\theta_t}
		\left(
			-g_R\frac{4 M_{BL} \mu v_u + g_{BL}^2 v_R^2 v_d}
			{2 M_{\tilde Y} \mu v_R^2}
			\epsilon_j
			-\frac{g_R g_{BL}^2}{2 M_{\tilde Y}} {v_L}_j^*
		\right)
	\right.
\\
	&
	\left.
		\quad - \frac{1}{3 \sqrt 2} g_{BL} s_{\theta_t}
		\left(
			g_{BL} \frac{g_R^2 v_R^2 v_d - 4 M_R \mu v_u}{2 M_{\tilde Y} \mu v_R^2} \epsilon_j
			+ \frac{g_{BL} g_R^2}{2 M_{\tilde Y}} {v_L}_j^*
		\right)
	\right.
\\
	&
	\left.
		\quad - Y_t c_{\theta_t}
		\left(
			\frac{M_{\tilde \gamma} v_R^2 v_d^2 + 4g_R^2 M_2 M_{BL} \mu v_d v_u}{4 M_{\tilde Y} M_2 v_R^2 \mu^2}
			\epsilon_j
			+ \frac{v_d M_{\tilde \gamma}}{4 M_{\tilde Y} M_2 \mu} {v_L}_j^*
		\right)
	\right]
\end{split}
\end{align}
\begin{align}
\begin{split}
	G^R_{\tilde t_1 t \nu_i}\approx
	\left(V_{PMNS}\right)_{ji}^*
	&
	\left[
		-\frac{1}{\sqrt 2} g_2 c_{\theta_t}
		\left(
			\frac{g_2 v_d}{2 M_2 \mu} \epsilon_j^*
			+ \frac{g_2}{2 M_2} {v_L}_j
		\right)
	\right.
\\
	&
	\left.
		\quad + \frac{1}{3 \sqrt 2} g_{BL} c_{\theta_t}
		\left(
			g_{BL} \frac{g_R^2 v_R^2 v_d - 4 M_R \mu v_u}{2 M_{\tilde Y} \mu v_R^2} \epsilon_j^*
			+ \frac{g_{BL} g_R^2}{2 M_{\tilde Y}} {v_L}_j
		\right)
	\right.
\\
	&
	\left.
		\quad - Y_t s_{\theta_t}
		\left(
			\frac{M_{\tilde \gamma} v_R^2 v_d^2 + 4g_R^2 M_2 M_{BL} \mu v_d v_u}{4 M_{\tilde Y} M_2 v_R^2 \mu^2}
			\epsilon_j^*
			+ \frac{v_d M_{\tilde \gamma}}{4 M_{\tilde Y} M_2 \mu} {v_L}_j
		\right)
	\right].
\end{split}
\end{align}
For the lightest stop vertex $\tilde t_1 \, b \, \tilde \chi_n^-:$
\begin{equation}
	 \quad  g_{\tilde t_1 b  \chi_n^\pm} = G^L_{\tilde t_1 b  \chi_n^\pm} P_L + G^R_{\tilde t_1 b  \chi_n^\pm} P_R \ ,
\end{equation}
with
\begin{align}
	G^L_{\tilde t_1  b \chi_n^\pm} & = 	Y_b c_{\theta_t} \mathcal{U}_{n 2}^* \ ,
	\\
	G^R_{\tilde t_1  b  \chi_n^\pm} & = 	- \frac{1}{\sqrt 2} g_2 c_{\theta_t} \mathcal{V}_{n 1} + Y_t s_{\theta_t} \mathcal{V}_{n 2} \ .
\end{align}
For the charged lepton components of the physical charginos, $\chi^\pm_{2+i} = \ell_i$, these $G$ parameters are approximated as
\begin{align}
	G^L_{\tilde t_1  b  \ell_i} & \approx
		Y_b c_{\theta_t}
		\frac{1}{\mu} \epsilon_i
	\\
	G^R_{\tilde t_1  b  \ell_i} & \approx 
	Y_t s_{\theta_t} \frac{m_{\ell_i}}{\sqrt 2 v_d \mu} {v_L}_i^* \ .
\end{align}

The approximations show that the top--neutrino channel is suppressed either by factors of $v_{d,u}/m_\text{soft}$ or by ${v_L}_i$ compared to the bottom-charged lepton channel. Therefore, the bottom-charged lepton channel dominates except for the case were the stop is mostly right-handed.
\subsection{Sbottoms}

For the lightest sbottom vertex $\tilde b_1 \, b \, \tilde \chi_{n}^0:$
\begin{equation}
	 \quad g_{\tilde b_1 b \chi_n^0} = G^L_{\tilde b_1 b \chi_n^0} P_L + G^R_{\tilde b_1 b \chi_n^0} P_R \ ,
\end{equation}
where $n$ labels the combined neutralinos (charginos) and neutrinos (charged leptons), and
\begin{align}
	G^L_{\tilde b_1 b \chi_n^0} & = 	-\frac{1}{\sqrt 2} g_R s_{\theta_b} \mathcal{N}_{n 1}^*
							+ \frac{1}{3 \sqrt 2} g_{BL} s_{\theta_b} \mathcal{N}_{n 5}^*
							- Y_b c_{\theta_b} \mathcal{N}_{n 3}^* \ ,
	\\
	G^R_{\tilde b_1 b \chi_n^0} & = 	\frac{1}{\sqrt 2} g_2 c_{\theta_b} \mathcal{N}_{n 2}
								- \frac{1}{3 \sqrt 2} g_{BL} c_{\theta_b} \mathcal{N}_{n 5}
								-Y_b s_{\theta_b} \mathcal{N}_{n 3} \ .
\end{align}
 For the neutrino components of the physical neutralinos, $\chi_{6+i} = \nu_i$, these $G$ parameters are approximated by
\begin{align}
	G^L_{\tilde b_1 b \nu_i} & \approx  {V_\text{PMNS}}_{ji} Y_b c_{\theta_b} \frac{\epsilon_j^*}{\mu} \ ,
	\\
	G^R_{\tilde b_1 b \nu_i} & \approx 	{V_\text{PMNS}}_{ji}^* Y_b s_{\theta_b} \frac{\epsilon_i}{\mu} \ .
\end{align}
For the lightest sbottom vertex $\tilde b_1 \, t \, \tilde \chi_n^-:$
\begin{equation}
	 \quad  g_{\tilde b_1 t \tilde \chi_n^-} = G^L_{\tilde b_1 t \tilde \chi_n^-} P_L + G^R_{\tilde b_1 t \tilde \chi_n^-} P_R \ ,
\end{equation}
with
\begin{align}
	G^L_{\tilde b_1  t \tilde \chi_n^\pm} & = 	Y_t c_{\theta_b} V_{n 2}^* \ ,
	\\
	G^R_{\tilde b_1  t \tilde \chi_n^\pm} & = 	-g_2 c_{\theta_b} U_{n 1} + Y_b s_{\theta_b} U_{n 2} \ .
\end{align}
For the charged lepton components of the physical charginos, $\chi^\pm_{2+i} = \ell_i$, these $G$ parameters are approximated as
\begin{align}
	G^L_{\tilde b_1  t \ell_i} & \approx 	Y_t c_{\theta_b} \frac{m_{\ell_i}}{v_d \mu} {v_L}_i \ ,
	\\
	G^R_{\tilde b_1  t \ell_i} & \approx 	Y_b s_{\theta_b} \frac{\epsilon_i^*}{\mu} \ .
\end{align}
In the sbottom sector, the bottom--neutrino and top--charged lepton channels are both unsuppressed except in the case of the mostly left-handed sbottom in which case the bottom--neutrino channel dominates.

\end{document}